\documentclass[fleqn]{2023SCGE}
\usepackage{microtype}
\usepackage{amsmath}

\usepackage{aas_macros}

\usepackage{amssymb,amsmath}
\usepackage{xspace}
\usepackage{multirow,array}
\usepackage[percent]{overpic}

\usepackage{enumitem}
\usepackage{xcolor}

\graphicspath{{./}{figures/}{figures/extra/}}

\newcommand{\hbt}{\textsc{hbt+}\xspace}
\newcommand{\hbtherons}{\textsc{hbt-herons}\xspace}
\newcommand{\msunh}{\mathrm{M}_\odot\;h^{-1}}

\newcommand{\rev}[1]{\textcolor{black}{#1}}

\begin{document}

\ensubject{subject}

\ArticleType{Article}
\SpecialTopic{SPECIAL TOPIC: }
\Year{2026}
\Month{April}
\Vol{66}
\No{1}
\DOI{??}
\ArtNo{000000}
\ReceiveDate{--}
\AcceptDate{--}

\title{\rev{The Sinking Statistics of Dark Matter Subhalos Across Hierarchical Levels}}

\author[1,2]{Wenkang Jiang}{}
\author[1,2]{Jiaxin Han}{{jiaxin.han@sjtu.edu.cn}}
\author[3]{Kun Xu}{}
\author[4]{Victor J. Forouhar Moreno}{}
\author[5]{Feihong He}{}
\author[6,7]{\\Zhaozhou Li}{}
\author[8]{Chunyan Jiang}{}
\author[1,2]{Yipeng Jing}{}
\author[1,2]{Xiaohu Yang}{}

\address[1]{Department of Astronomy, School of Physics and Astronomy, Shanghai Jiao Tong University, Shanghai 200240, China}
\address[2]{State Key Laboratory of Dark Matter Physics, School of Physics and Astronomy, Shanghai Jiao Tong University, Shanghai 200240, China}
\address[3]{Center for Particle Cosmology, Department of Physics and Astronomy, University of Pennsylvania, Philadelphia, PA 19104, USA}
\address[4]{Leiden Observatory, Leiden University, PO Box 9513, NL-2300 RA Leiden, the Netherlands}
\address[5]{Kavli Institute for Astronomy and Astrophysics, Peking University, Beijing 100871, China}
\address[6]{School of Astronomy and Space Science, Nanjing University, Nanjing 210093, China}
\address[7]{Key Laboratory of Modern Astronomy and Astrophysics, Nanjing University, Ministry of Education, Nanjing 210093, China}
\address[8]{Key Laboratory for Research in Galaxies and Cosmology of Chinese Academy of Sciences, Shanghai Astronomical Observatory, Nandan Road 80, \\Shanghai 200030, China}

\AuthorMark{Jiang W.}

\AuthorCitation{Jiang W., et al.}

\abstract{
\rev{We investigate the mergers among subhalos in a $\Lambda$CDM simulation, focusing on two fundamental aspects overlooked by previous studies: 1) how to identify mergers robustly; 2) the statistics of mergers across the subhalo hierarchy.} To this end, we make use of the \hbt subhalo finder that tracks subhalo evolution across hierarchy levels, identifying the coalescence of subhalo cores in phase space as
a ``sinking'' event. This coalescence marks a distinct stalled phase in orbital decay, providing a physically motivated and natural definition of a resolved merger. \rev{Moreover, the phase transition occurs over a very short timescale, making the statistics robust to numerical resolutions.} Our main findings are as follows. (1) More than 90\% of sinking events occur between adjacent subhalo levels, while cross--level pathways arise from tidal stripping, group accretion, and numerical constraints. (2) Resolved sinking events are predominantly major mergers (mass ratios $>1{:}10$), whereas the contribution from minor mergers decreases with the dynamical age of the host halo. (3) Although deep-level subhalos typically have low mass ratios relative to the host halo, their mass ratios relative to their direct parents are substantially larger, significantly enhancing their sinking probability. Consequently, the satellite--satellite sinking rate can rival or even exceed the central--satellite sinking rate at lower mass thresholds. (4) Satellite--satellite sinking events are spatially biased toward the outer regions of the host halo, suggesting that the central tidal field suppresses orbital decay between satellite systems.
}



\maketitle

\twocolumn

\section{Introduction}\label{sec:intro}
Dark matter subhalos are self-bound substructures embedded within larger dark matter halos, forming through hierarchical structure assembly in the $\Lambda$CDM paradigm~\cite{subhalo1,subhalo2,subhalo3,subhalo4}. \rev{After accretion, massive subhalos undergo orbital decay primarily driven by dynamical friction, spiraling inward toward the host center. This orbital decay can lead to subhalo mergers and, by extension, mergers of the galaxies they host}~\cite{friction_evidence_Ostriker,friction_evidence_Malumuth,friction_evidence_White,friction_evidence_Merritt}, thereby contributing to the growth of central galaxies. Hence, determining how rapidly and how often such subhalo mergers occur is a crucial step in modeling galaxy evolution.

A key quantity in this context is the subhalo merger timescale, which characterizes the time required for a newly accreted subhalo to \rev{complete its orbital decay and merge} with its host. Several numerical studies have derived fitting functions for this timescale using both idealized and cosmological simulations~\cite{merger_bookBinney,merger_Lacey93,merger_Boylan2008,merger_Jiang2008,merger_Xu2025}. These works consistently show that the \rev{accretion} redshift is an important factor, as it sets the dynamical time of the host halo at accretion: subhalos accreted at higher redshifts (earlier times) generally experience shorter dynamical times and thus merge more rapidly. Other critical factors established by these works include the orbital parameters (e.g., circularity) and the subhalo-to-host mass ratio at accretion: subhalos on more radial orbits or with higher mass ratios generally merge more rapidly.

Merger timescale prescriptions play a central role in semi-analytic models~\cite{SAM_LGALAXIES,SAM_GALFORM,SAM_GEAE,SAM_SHARK}, where they are used to translate the merger history of subhalos into predictions for galaxy-galaxy mergers. This connection enables SAMs to study a range of merger-driven physical processes, such as morphological transformations~\cite{morphology1,morphology_disk1,morphology_bulge2}, starbursts~\cite{merger_starburst1,merger_starburst2,merger_starburst3}, and AGN activity~\cite{merger_AGN1,merger_AGN2}.

The subhalo merger rate is another key quantity, describing the frequency of merger events in a population~\cite{merger_rate_Stewart,merger_rate_Wetzel,merger_rate_Angulo,merger_rate_Hopkins,merger_rate_Kitzbichler,merger_rate_Wang,merger_rate_Rodriguez}. While it is fundamentally defined as the number of mergers occurring per unit time per subhalo, the precise definition varies across studies. For instance, some works measure the rate per descendant halo~\cite{merger_rate_Stewart}, while others report it per progenitor subhalo~\cite{merger_rate_Wetzel}. Despite these methodological differences, estimates from different models are found to be broadly consistent, typically agreeing within a factor of two~\cite{review_Hopkins}.

A key complication in these studies is that there is no universal definition of what a subhalo ``merger'' is. The lack of a unique definition complicates the interpretation of simulation data and its application to galaxy formation models. Operational definitions range from the loss of the majority of a subhalo's mass~\cite{merger_rate_Stewart}, coalescence defined by multiple progenitor subhalos sharing a common descendant subhalo~\cite{merger_Jiang2008,merger_rate_Wetzel}, losing a given fraction of the initial specific orbital angular momentum~\cite{merger_Boylan2008,merger_Xu2025}, or reaching a certain physical separation~\cite{merger_Poulton}. To further complicate matters, the choice of the (arbitrary) threshold used within these definitions can modify the measured merger rate. For instance, J. M. Solanes et al.~\cite{merger_Solanes} demonstrated that defining subhalo mergers based on a threshold of 5\% of the initial specific angular momentum leads to the premature termination of tracked orbital decay. The variety of merger definitions in use reflects our insufficient understanding of a core question regarding the process: what defines a merger physically? Addressing this question can provide a more self-consistent and physically motivated definition of merger events.

Another complication stems from the hierarchical nature of accretion itself. It is common for subhalos to be accreted as members of groups (``\rev{group accretion}''), which naturally establishes a nested, multi-level hierarchy within the host halo~\cite{group_infall1,group_infall2,group_infall3,group_infall4,group_infall5,group_infall6}. This hierarchy entails that merging with the central subhalo is not the only fate for accreted subhalos, and some will merge with other satellite subhalos during their orbital evolution. \rev{Although central--satellite and satellite--satellite mergers are often discussed as separate channels, the physical pathways and final fates of subhalos in multi-level systems remain poorly understood.}

\rev{We study subhalo mergers associated with dynamical-friction-driven orbital decay using a dark-matter-only (DMO) cosmological simulation analyzed with the {\hbt} subhalo finder~\cite{Han12,Han18}. The {\hbt} sinking criterion identifies a merger when a subhalo approaches the resolved core scale of its host subhalo and becomes indistinguishable from that core in phase space, marking the final stage of orbital decay that can be resolved in the simulation.}

\rev{We define a ``sinking event'' as the first snapshot at which a subhalo satisfies this criterion with respect to another subhalo. In the {\hbt} catalog, the subhalo is then treated as merged into that host. Throughout this paper, ``sinking'' refers specifically to this event-based merger definition rather than the full orbital-decay process. Accordingly, we analyse the resulting population of sunken subhalos, as well as their sinking rates and spatial distributions. This work aims to:}

\begin{enumerate}[label=(\roman*)]
\item \rev{Investigate the physical processes experienced by dark matter subhalos during the late stages of orbital decay in the simulation.}

\item Characterize the final fates of sunken subhalos within multi-level hierarchical systems.

\item \rev{Investigate the statistical trends of sinking rates and locations, and their dependence on redshift and host halo mass.}
\end{enumerate}
The paper is organized as follows. \rev{Section~\ref{sec:method} describes the simulation and the {\hbt} algorithm, as well as the selection criteria used to define the sunken subhalo sample analyzed in this work.} Section~\ref{sec:sink_fate} displays the fates of sunken subhalos within the multi-level hierarchy and classifies the cross‑level sinking pathways. Section~\ref{sec:merger_rate} introduces two complementary definitions of the sinking rate—group‑level and halo‑level—and presents their dependence on mass ratio, redshift, and host halo mass. Section~\ref{sec:spatial_distribution} examines the radial distribution of sinking events and decomposes it by sinking type. \rev{Section~\ref{sec:robustness} examines the resolution dependence of the sinking criterion and tests the robustness of the prior statistical results using a higher-resolution simulation. Section~\ref{sec:disucssion} compares the original {\hbt} criterion with the bidirectional criterion adopted in \textsc{HBT-HERONS}\xspace \cite{Victor}, and discusses the astrophysical implications of the sinking statistics presented in this work.} Section~\ref{sec:summary} summarizes our main findings.

\section{Methods}
\label{sec:method}
\subsection{Simulation \& Subhalo Finder}
Our analysis is based on a dark-matter-only simulation from the CosmicGrowth suite~\cite{CosmicGrowth}, specifically the L600 run. The simulation adopts a WMAP cosmology with parameters $H=71\,\mathrm{km \, s^{-1}\, Mpc^{-1}}$, $\Omega_{M_0}=0.268$ and $\Omega_{\Lambda}=0.732$, and follows the evolution of $3072^3$ particles within a periodic box of side length $600\,\mathrm{Mpc}\,h^{-1}$. The particle mass is $m_p \approx 5.5 \times 10^8\,\mathrm{M}_\odot\,h^{-1}$. The gravitational force is softened with a Plummer-equivalent softening length of $\epsilon = 3.3\,\mathrm{kpc}\,h^{-1}$ in comoving coordinates. In this work, halo masses are defined as the mass enclosed within the radius where its inner matter density equals the virial density predicted from the spherical collapse model, $\rho_{\rm vir} = \Delta_{\rm vir}\rho_{\rm crit}$, where $\Delta_{\rm vir}=18\pi^2+82[\Omega_{\rm m}(z)-1]-39[\Omega_{\rm m}(z)-1]^2$~\cite{virial_overdensity}.

We identify subhalos and construct their merger trees using the {\hbt} (Hierarchical Bound-Tracing) algorithm~\cite{Han18}. In {\hbt}, the tracking algorithm operates in two distinct regimes: (1) when a subhalo is the central structure of its own host halo, its bound particles are continuously updated using the particle reservoir of the corresponding FoF group; (2) once it falls into a more-massive host and becomes a satellite, {\hbt} records its particle membership at the accretion time and subsequently removes particles that become unbound during its orbital evolution. A subhalo is considered resolved and is tracked as a self-bound object as long as it retains at least $N_{\mathrm{min}} = 20$ bound particles.
If the number of bound particles of a subhalo falls below this threshold, the subhalo is flagged as unresolved. {\hbt} subsequently tracks it as an ``orphan'' subhalo, following only the position and velocity of the most bound particle at the last resolved snapshot. \rev{We clarify the distinction between ``orphan'' and ``tidal disruption'' in this work. Here, ``orphan'' refers strictly to a subhalo that is no longer resolved by the {\hbt} tracking algorithm. We do not attempt to infer physical disruption below the resolution limit; therefore, we use ``tidal disruption'' operationally to also include cases in which the subhalo mass falls below the resolution limit.}

{\hbt} provides a complete and self-consistent record of the hierarchical relationships between subhalos, tracing their origin to \rev{group accretion} events. Within this structure, the hierarchical level (denoted by the symbol $\ell$ hereafter) of a subhalo is quantified by the number of distinct host subhalos that sequentially contain the subhalo. Specifically, the central subhalo is designated as $\ell=0$. Based on the hierarchical level, we define the following terminology to describe hierarchical relationships:

\textbf{Child Subhalo}:  A subhalo (of level $\ell$) is considered the child of the specific subhalo of level $\ell-1$ or shallower levels, in which it is sequentially embedded.

\textbf{Direct Parent Subhalo}:  The immediate host subhalo within which a child subhalo is directly embedded. This corresponds to the host subhalo at the next shallower hierarchical level.

\textbf{Parent Subhalo}: The set of all hosts that contain the child subhalo at successively shallower hierarchical levels. This includes its direct parent and all ancestors up to the central subhalo.

\begin{figure}[t]
    \centering
    \includegraphics[width=0.5\textwidth]{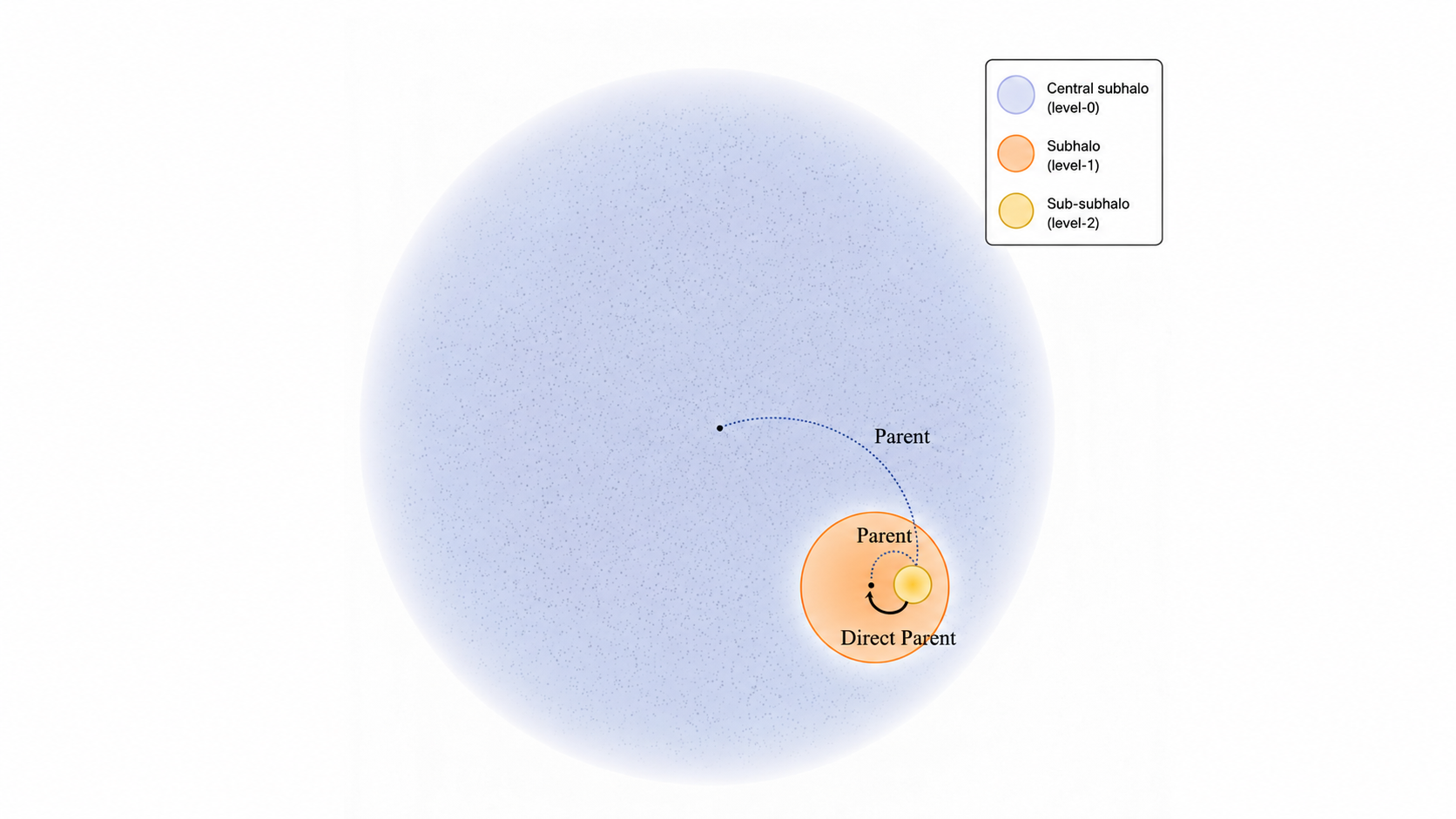}
    \caption{\rev{Illustration of a hierarchical halo system consisting of a central subhalo (purple) and a subhalo group with level-1 (orange) and level-2 (yellow) subhalos. The level-2 subhalo resides within the level-1 subhalo (its direct parent), which in turn is embedded in the central subhalo.}}
    \label{fig:illustration_hierarchy}
\end{figure}

\rev{Figure~\ref{fig:illustration_hierarchy} illustrates the position of a level-2 subhalo within the host-halo subhalo hierarchy.
In this example, the level-2 subhalo is embedded within a level-1 subhalo (its direct parent), which in turn resides within the central subhalo of the host halo.} 

\subsection{Identifying Sunken Subhalos}\label{sub_sec:sink_criterion}

\begin{figure}[t]
    \centering
    \includegraphics[width=0.45\textwidth]{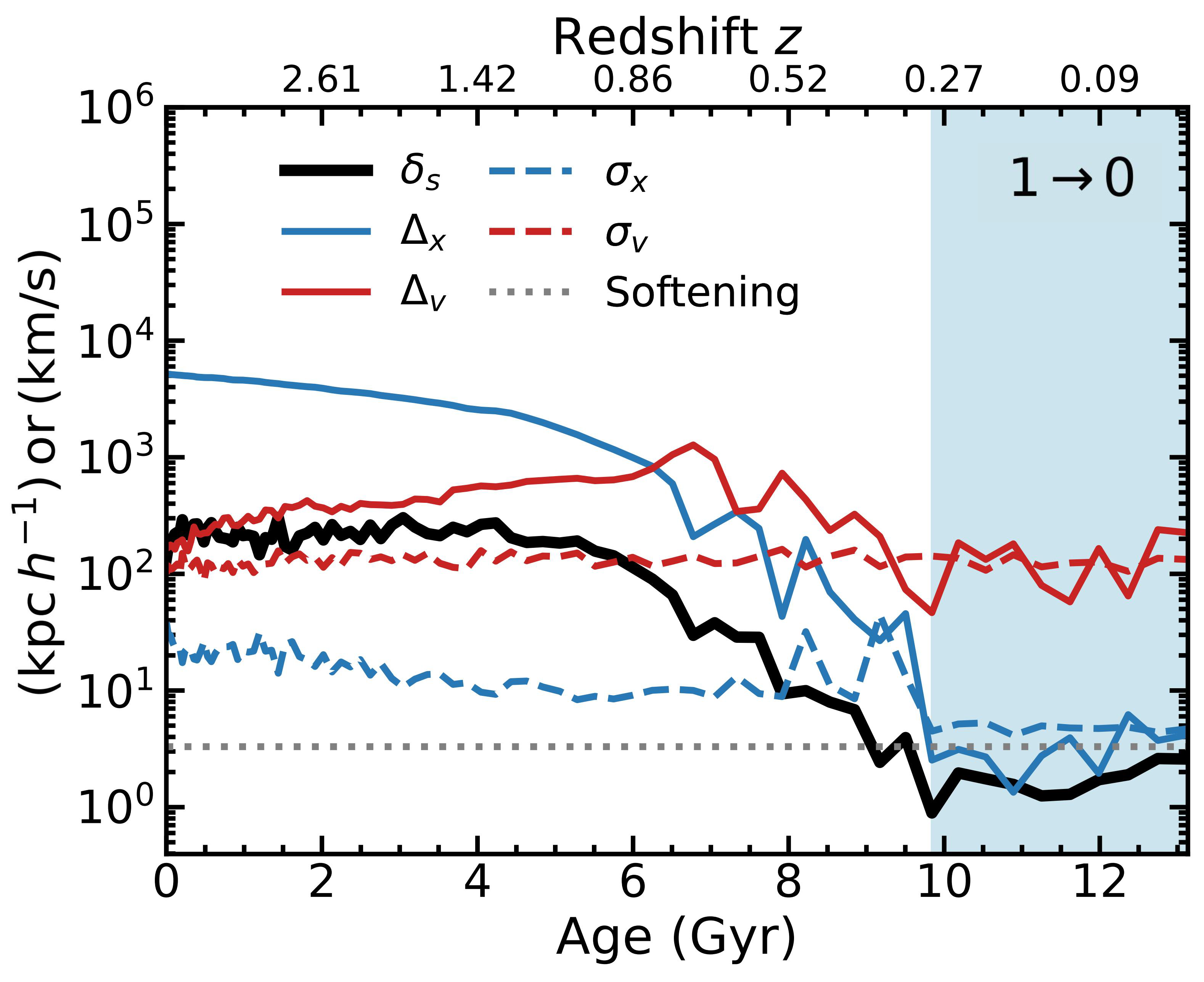}
    \caption{A resolved sinking event between two subhalos. The figure shows the evolution of the kinematic separation between a level-1 subhalo and its central subhalo. This level-1 subhalo was accreted into its host halo at redshift \rev{$z=1.15$} with a peak mass ratio of $m_{\rm peak}/M_{\rm vir}=0.35$. We also present the spatial and velocity dispersions of the parent subhalo ($\sigma_x$ and $\sigma_v$), along with the instantaneous spatial separation ($\Delta_x$) and relative velocity ($\Delta_v$) between the two subhalos. The blue shaded region marks the post-sinking stage after the criterion is first satisfied.}
    \label{fig:ds_evolution}
\end{figure}

\rev{Subhalos undergoing orbital decay within a host halo gradually lose orbital energy and angular momentum due to dynamical friction, causing their orbits to shrink toward the central region of the host subhalo. When approaching the characteristic core scale of the host, further orbital decay becomes difficult to resolve at finite resolution, as the subhalo instead remains confined to the vicinity of the host core scale, making it effectively indistinguishable from the host in phase space. In {\hbt}, this regime is therefore identified as the final resolvable stage of orbital decay, and we refer to the corresponding transition as a sinking event.}

\rev{To quantify this transition, {\hbt} compares the phase-space dispersion of a child subhalo to that of a random particle in the very center of the host. 
Specifically, the spatial ($\sigma_x$) and velocity ($\sigma_v$) dispersions of the 20 most bound particles of the host are used to characterize the local phase-space scale near the center. The normalized phase-space distance between a child subhalo and its parent is then defined as}

\begin{equation}
\delta_{\rm s} = \frac{|\mathbf{x_{\rm c}} - \mathbf{x}_{\rm p}|}{\sigma_x} + \frac{|\mathbf{v_{\rm c}} - \mathbf{v_{\rm p}}|}{\sigma_v},
\label{eq:orig_sink_algorithm}
\end{equation}
where $\mathbf{x_{\rm c}}$ and $\mathbf{v_{\rm c}}$ are the position and velocity of the most bound particle of the child, while $\mathbf{x_{\rm p}}$ and $\mathbf{v_{\rm p}}$ are the averaged position and velocity of the 20 most bound particles of the parent. A sinking event is flagged when $\delta_{\rm s}$ falls below a predefined threshold (set to 2 in this work), indicating the subhalo has become indistinguishable from a random particle in the core of the parent.

This check is performed iteratively at each snapshot for every subhalo, tracing its lineage upward from its direct parent toward the central subhalo of the host halo. \rev{By construction, the algorithm searches for sinking events along parent-child links in the hierarchy} 
rather than between peers at the same hierarchical level or unrelated substructures within the same host.

\rev{Figure~\ref{fig:ds_evolution} shows the evolution of a representative level-1 subhalo whose phase-space distance with respect to its direct parent eventually drops below the threshold. Consistent with the example presented by J. Han et al.~\cite{Han18}, the evolution exhibits two distinct phases. Following accretion, the subhalo undergoes orbital decay driven by dynamical friction. Both its spatial and velocity separations from the central subhalo decrease steadily. Once these separations become comparable to the characteristic core scales of the central subhalo ($\sigma_x$ and $\sigma_v$), the inspiral stalls and the subhalo begins to oscillate around the parent core. Such stalling behavior has been reported in previous studies and may arise from core dynamics, particle resonances, or discreteness effects ~\cite{core_stalling1,core_stalling2,core_stalling3,core_stalling4}.}

\rev{The original {\hbt} implementation evaluates the approach of a child subhalo to the core of its parent. A bidirectional variant, used in {\hbtherons} \cite{Victor}, also tests the reverse distance from the parent to the child's core. We construct our main sample with the original single-sided criterion, and examine the additional reverse-link events separately in Section~\ref{sec:reverse-link}. We treat these reverse-link events as a separate comparison sample, as they arise from a distinct mechanism during late-stage evolution and exhibit different subsequent structural evolution.}

\begin{figure}[t]
    \centering
    \includegraphics[width=.9\linewidth]{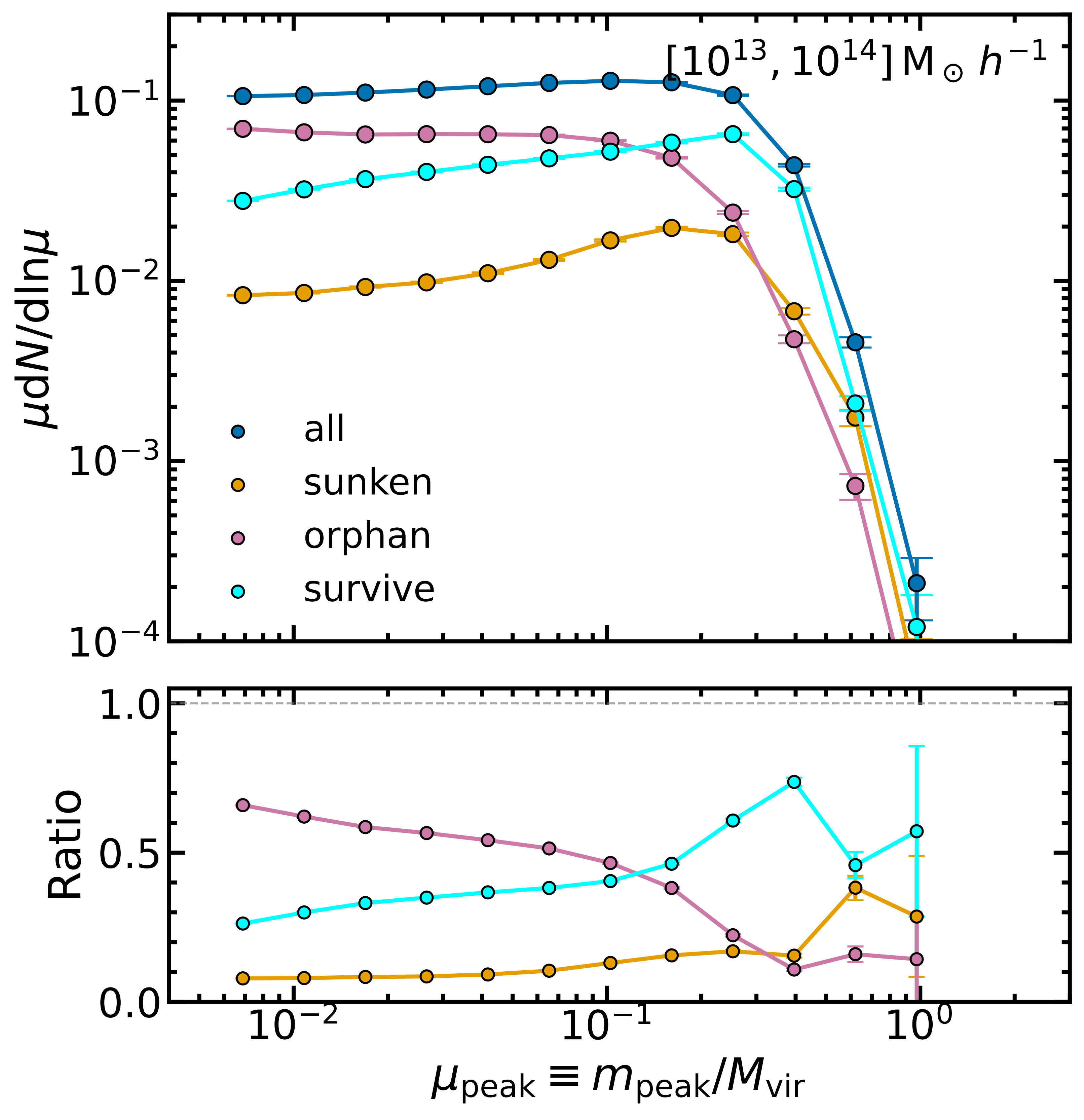} 
    \caption{The subhalo peak mass function (PMF) for host halos with $M_{\rm vir}$ in the range $[10^{13},10^{14}]\,{\rm M_\odot}\,h^{-1}$ at $z=0$, including satellite subhalos of all hierarchical levels. The total PMF is decomposed into the surviving, orphan, and sunken populations, each shown as a solid line in a different color. The lower panel shows the ratio of the PMF of each population to that of the overall population.}
    \label{fig:PMF_DECOMPOSE}
\end{figure}

\begin{figure*}
    \centering
    \includegraphics[width=0.8\linewidth]{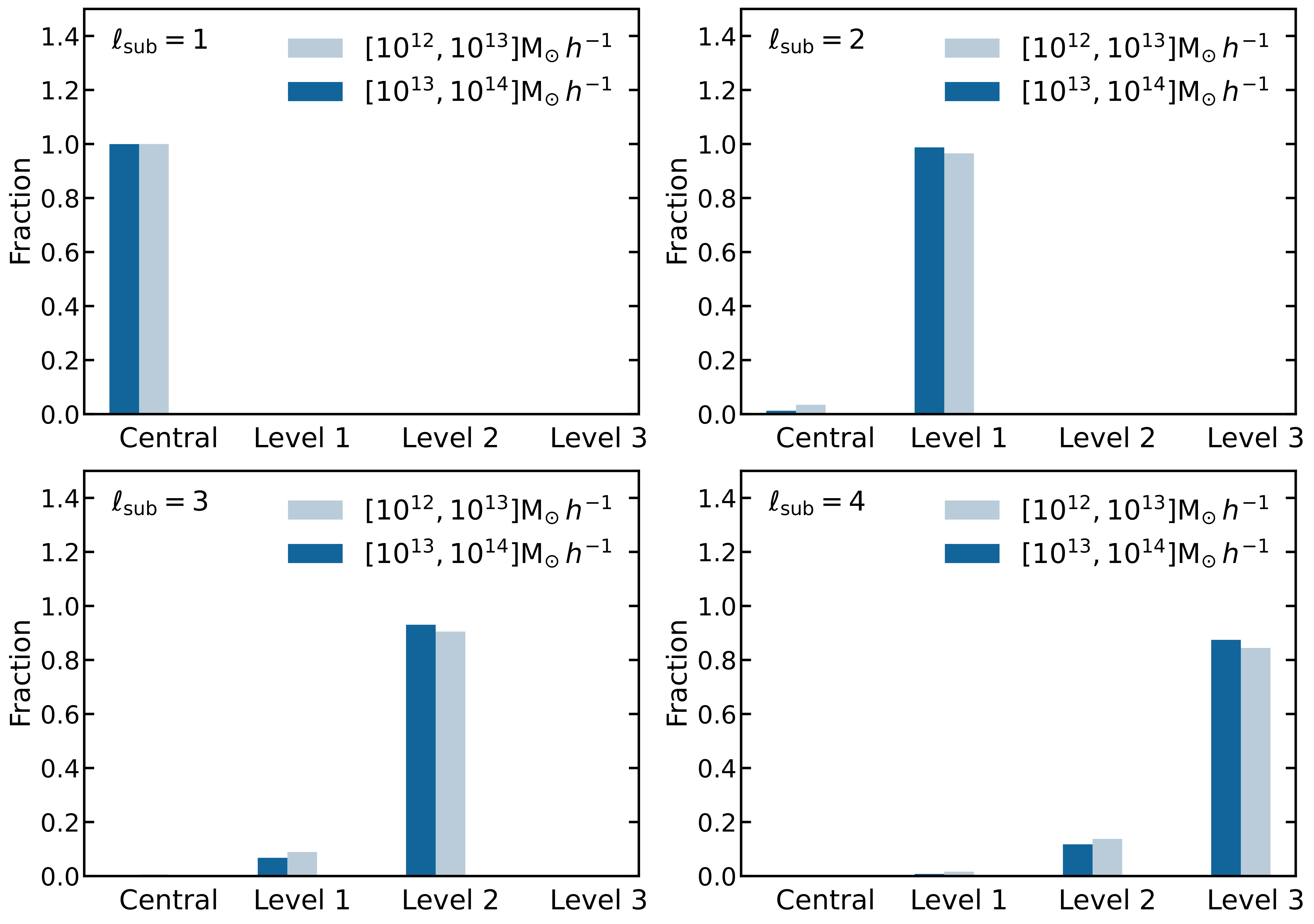}
    \caption{Distributions of the levels of the final parent subhalo associated with the sinking event for child subhalos at a given level. The four panels correspond to child subhalos of the first four levels. Dark blue bars denote results in more massive halos, while the light blue bars denote those in less massive halos.}
    \label{fig:sink_parent_level}
\end{figure*}

\subsection{Subhalo Samples}

\rev{We now define the subset of sinking events used for the statistical analysis. In the {\hbt} catalog, the sinking criterion can still be evaluated after a subhalo becomes an ``orphan''. Such post-orphan trajectories are not suitable for our physical interpretation of sinking, because an orphan no longer represents a resolved massive perturber capable of experiencing dynamical-friction-driven orbital decay in the usual sense. We therefore construct the main statistical sample from events for which both the child and parent remain resolved when the criterion is first satisfied.}

More specifically, we impose a resolution cut on subhalos involved in sinking events,
\begin{equation}
m_{\rm sink} \geq 20\,m_{\rm p},
\label{eq:sink_crit}
\end{equation}
and require that selected subhalos are well resolved at the time of accretion using a peak-mass threshold,
\begin{equation}
m_{\rm peak} \geq 100\,m_{\rm p}.
\label{eq:peak_crit}
\end{equation}

After applying these criteria, we identify a total of 357379 sinking events occurring in host halos of virial mass $[10^{12},10^{13}]\mathrm{M}_\odot\,h^{-1}$ across all snapshots, and 61870 events in hosts of mass $[10^{13},10^{14}]\mathrm{M}_\odot\,h^{-1}$. This sample forms the basis for our subsequent statistical analysis. 

\rev{Figure~\ref{fig:PMF_DECOMPOSE} illustrates the composition of the total subhalo population in terms of the surviving, orphan, and sunken populations. The relative contribution from surviving subhalos is around $50\%$ at the high $\mu_{\rm peak}$ end, consistent with the expectation from the unified subhalo distribution model of \cite{Han12}. Towards the low $\mu_{\rm peak}$ end, this fraction decreases due to the finite mass resolution of the simulation such that more and more surviving subhalos become unresolved and are counted as orphans. The sunken population is more abundant than the orphan population for peak mass ratios above $\mu_{\rm peak}\gtrsim3\times10^{-1}$. At the low mass end ($\mu_{\rm peak}\lesssim 10^{-1}$), however, they contribute only $\sim 10\%$ to the total peak mass function.} 

\section{Hierarchical Sinking Pathways of Subhalos}
\label{sec:sink_fate}

The prevailing framework in the literature classifies subhalo mergers into two broad channels: central--satellite and satellite--satellite mergers. While conceptually clear, this binary scheme does not fully capture the complexity inherent in hierarchically structured systems. \rev{For the sinking events identified by the {\hbt} criterion, both channels encompass diverse sinking pathways in hierarchically structured systems. We therefore classify each sinking event by the level of the sunken subhalo and the level of its final parent.}

\begin{figure}[t]
    \centering
    \includegraphics[width=.9\linewidth]{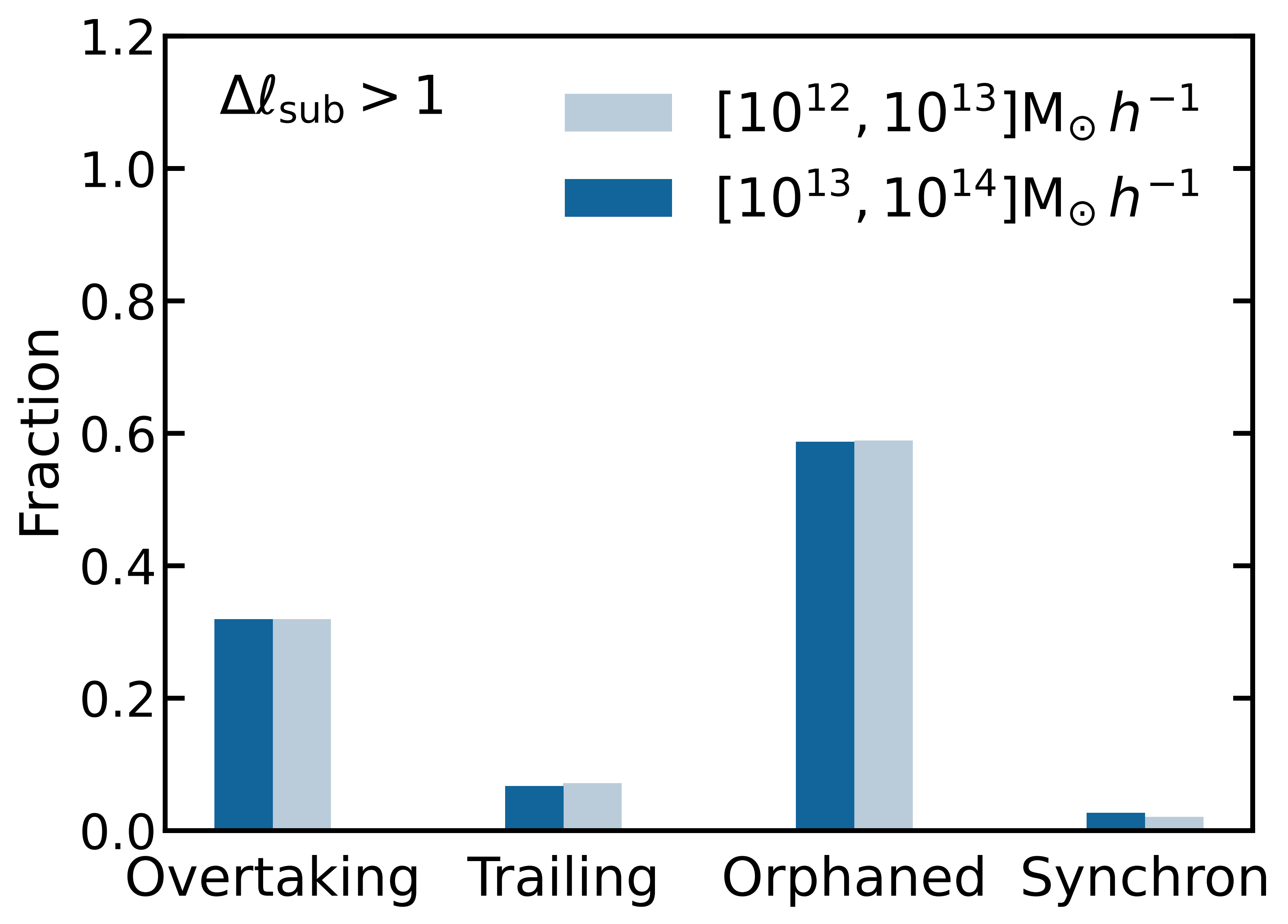}
    \caption{ The relative abundance of the \rev{four cross--level pathways}. The plot shows the distribution for all of the deep-level ($\ell>1$) sinking events across all redshifts. We use bars with different colors to indicate the \rev{events in each cross--level pathway} measured from different host halo mass ranges.}
    \label{fig:sink_parent_diff_scenarios}
\end{figure}

Figure~\ref{fig:sink_parent_level} shows the distribution of parent levels for sunken child subhalos binned by their own level $\ell_{\rm sub}$. Our analysis reveals a clear hierarchical sinking pattern. The vast majority of sinking events occur between a subhalo and its direct parent (level $\ell_{\rm sub}-1$). This is consistent with the findings in R. E. Angulo et al.~\cite{merger_rate_Angulo}, where they demonstrated that subhalos involved in satellite--satellite mergers are dynamically associated with each other prior to accretion, rather than forming through random encounters within the dark matter halo.

However, \rev{events following cross--level pathways} also constitute a non-negligible fraction of sinking events, and this fraction increases slightly with $\ell_{\rm sub}$. For these cross--level events, the level difference between the child and its final parent is predominantly 2. Notably, we find that only a very small fraction ($\approx 0.5\%$) of sinking events occur directly between the central subhalos (level 0) and child subhalos with $\ell_{\rm sub} > 2$. This indicates that nearly all the central--satellite mergers are contributed by the subhalos and sub-subhalos in the simulation. Furthermore, we find that this parent-level distribution exhibits little dependence on the mass of the main host halo, indicating that the pathway statistics are governed by local hierarchy rather than global halo properties.

\rev{For cross--level sinking events, the child does not sink into its direct parent, but instead ends up sinking into a shallower-level parent. We classify these events according to the relative timing of the child and its direct parent:}
\begin{itemize}

\item \rev{\textit{Overtaking Pathway:} The child subhalo sinks into a shallower-level parent before its direct parent does. In other words, the child bypasses its direct parent and reaches the larger system first. This can occur when tidal stripping or dynamical heating weakens the binding between the child and its direct parent, allowing the child to evolve independently.}

\item \rev{\textit{Trailing Pathway:} The direct parent sinks first, while the child remains resolved for some additional time and later sinks into the shallower-level parent. This corresponds to a trailing case in which the child follows the sinking of its direct parent.}

\item \rev{\textit{Orphaned Pathway:} The direct parent becomes unresolved before the child sinks. The child then survives as a separate object and later sinks into a shallower-level parent. This pathway is therefore sensitive to mass resolution, since a higher-resolution simulation may keep the original parent resolved for a longer time.}

\item \rev{\textit{Synchronous Pathway:} The child and its direct parent sink into a shallower-level parent in the same snapshot. This means that the two sinking events are simultaneous within the time resolution of the simulation outputs. With finer snapshot sampling, some of these events might be separated into overtaking or trailing cases.}

\end{itemize}

Subsequently, we measure the relative abundance of these four pathways among all \rev{events following cross--level pathways}. The results are shown in Figure~\ref{fig:sink_parent_diff_scenarios}. \textit{\rev{Orphaned Pathway}} and \textit{\rev{Overtaking Pathway}} are the two dominant pathways among the \rev{cross--level events}. The high fraction of \textit{\rev{Orphaned Pathway}} suggests that insufficient particle resolution plays a significant role in facilitating cross--level events.
The prevalence of \textit{\rev{Overtaking Pathway}} indicates that tidal stripping of the nearby massive subhalos can efficiently unbind a subhalo from its direct parent, allowing it to sink independently into another one. It highlights the complex interplay of physical stripping and numerical effects in determining the fate of group-accreted subhalos~\cite{ejected_ludlow,ejected_sales,ejected_Wang,vdb_dissecting}.

Our analysis of several \rev{pathways} has important applications to semi-analytic models. In some models, when a subhalo group is accreted, the merger clocks of its satellites are reset based on the new host halo properties~\cite{group_merger_recalculate1,group_merger_recalculate2}. Our findings offer a more nuanced view. The prevalence of the \rev{next-level pathway} supports the idea that many satellites remain bound to their group halo, for which maintaining the original merger timescale may be appropriate~\cite{group_merger_recalculate3,JiangCY_2010}.

\section{The Sinking Rate}
\label{sec:merger_rate}

The merger rates between subhalos and galaxies have also been extensively studied in the literature \cite{review_Hopkins,merger_rate_Kitzbichler,merger_rate_Hopkins,merger_rate_Stewart,merger_rate_Wang,merger_rate_Wetzel,merger_rate_Rodriguez}. In this section, we examine \rev{the sinking rate of subhalos, with} a finer classification based on the hierarchical level $\ell$ of satellites.
\rev{For the rate measurements, we focus on direct parent–child sinking events, which account for more than 90\% of the sample.
} We compare the sinking rates across different groups and halos and find that satellite--satellite mergers constitute a non-negligible fraction of the overall sunken subhalo population.

\subsection{Two Complementary Definitions of the Sinking Rate}
Our finding that identified sinking events occur predominantly between a child and its direct parent motivates two complementary definitions:

\textbf{1. Group-level sinking rate.}
A direct parent subhalo of level $\ell-1$ and its children of level $\ell$ can be defined as a subhalo group of $\ell-1$ within a dark matter halo. Investigating the sinking rate within each such group provides a quantitative measure of sinking rate across hierarchical levels and reveals how the efficiency of internal orbital decay within a group depends on its hierarchical level. Here, we define the group-level sinking rate for subhalos of level $\ell$ as:
\begin{equation}
f_{\mathrm{group},\ell}(\mu_{\rm rel} \mid m_\mathrm{parent},z) =
\frac{
\Delta N_{\mathrm{sink},\ell}(\mu_{\rm rel} \mid m_\mathrm{parent}, z)
}{
N_{\rm group,\ell-1}\,\Delta \tau\,\Delta \ln \mu_{\rm rel}
},
\end{equation}
where $\mu_{\rm rel}\equiv m_{\rm child}/m_{\rm parent}$ is the ratio of the child subhalo’s peak mass to the peak mass of the descendant of its direct parent after \rev{sinking}. We define $\Delta \tau \equiv \Delta t / t_{\rm dyn}(z)$, where $\Delta t$ is the time interval between two snapshots and $t_{\rm dyn}(z)$ is the dynamical timescale at $R_{\rm vir}$ at that redshift. The denominator is constructed as follows: at a given simulation snapshot (redshift $z$), we select all level-$(\ell-1)$ groups whose group mass—defined as the peak mass of the direct parent subhalo at that snapshot—falls within a specified range. $N_{\rm group,\ell-1}$ is the total number of such groups. The numerator $\Delta N_{{\rm sink},\ell}(\mu_{\rm rel} \mid m_\mathrm{parent}, z)$ counts, within the same selected groups, all events occurring over the interval $\Delta \tau$ where a level-$\ell$ subhalo sinks into its direct level-$(\ell-1)$ parent, with a mass ratio residing within  $[\ln \mu_{\rm rel},\ln \mu_{\rm rel}+\Delta \ln \mu_{\rm rel}]$. This measures the sinking rate per subhalo group per unit time. While conceptually similar to A. R. Wetzel et al.~\cite{merger_rate_Wetzel}, our key difference is the dissection of the rate by subhalo level $\ell$, allowing us to isolate distinct sinking pathways (e.g., level-2 into level-1 vs. level-1 into the central subhalo) that are often \rev{combined}
in previous studies.

\textbf{2. Halo-level sinking rate.}
This perspective aggregates all sinking events of level-$\ell$ subhalos within an entire host halo. We define the halo-level sinking rate for subhalos of level $\ell$ as:
\begin{equation}
f_{\rm halo,\ell}(\mu_{\rm peak} \mid M_\mathrm{vir},z) =
\frac{
\Delta N_{{\rm sink},\ell}(\mu_{\rm peak} \mid M_{\rm vir}, z)
}{
N_{\rm halo}\,\Delta \tau\,\Delta \ln \mu_{\rm peak}
},
\end{equation}
where $\mu_{\rm peak} \equiv m_{\rm peak}/M_{\mathrm{vir}}$ is the ratio of the subhalo's peak mass to the host halo's virial mass. $N_{\rm halo}$ is the number of host halos in the specified virial mass range at redshift $z$. The numerator $\Delta N_{\mathrm{sink},\ell}$ counts, over the time interval $\Delta \tau$ between snapshots, all level-$\ell$ child subhalos that sink into their direct parents within those host halos with peak mass ratios residing within $[\ln \mu_{\rm peak},\ln \mu_{\rm peak}+\Delta \ln \mu_{\rm peak}]$. This measures the total number of level-$\ell$ sunken subhalos per host halo per unit time, matching conventions such as R. E. Angulo et al.~\cite{merger_rate_Angulo}. Finally, to capture sinking events occurring in the outskirts, we include all sinking events within subhalo groups defined via the host halo’s FoF membership.

\begin{figure}[!t]
    \centering
    \includegraphics[width=.9\linewidth]{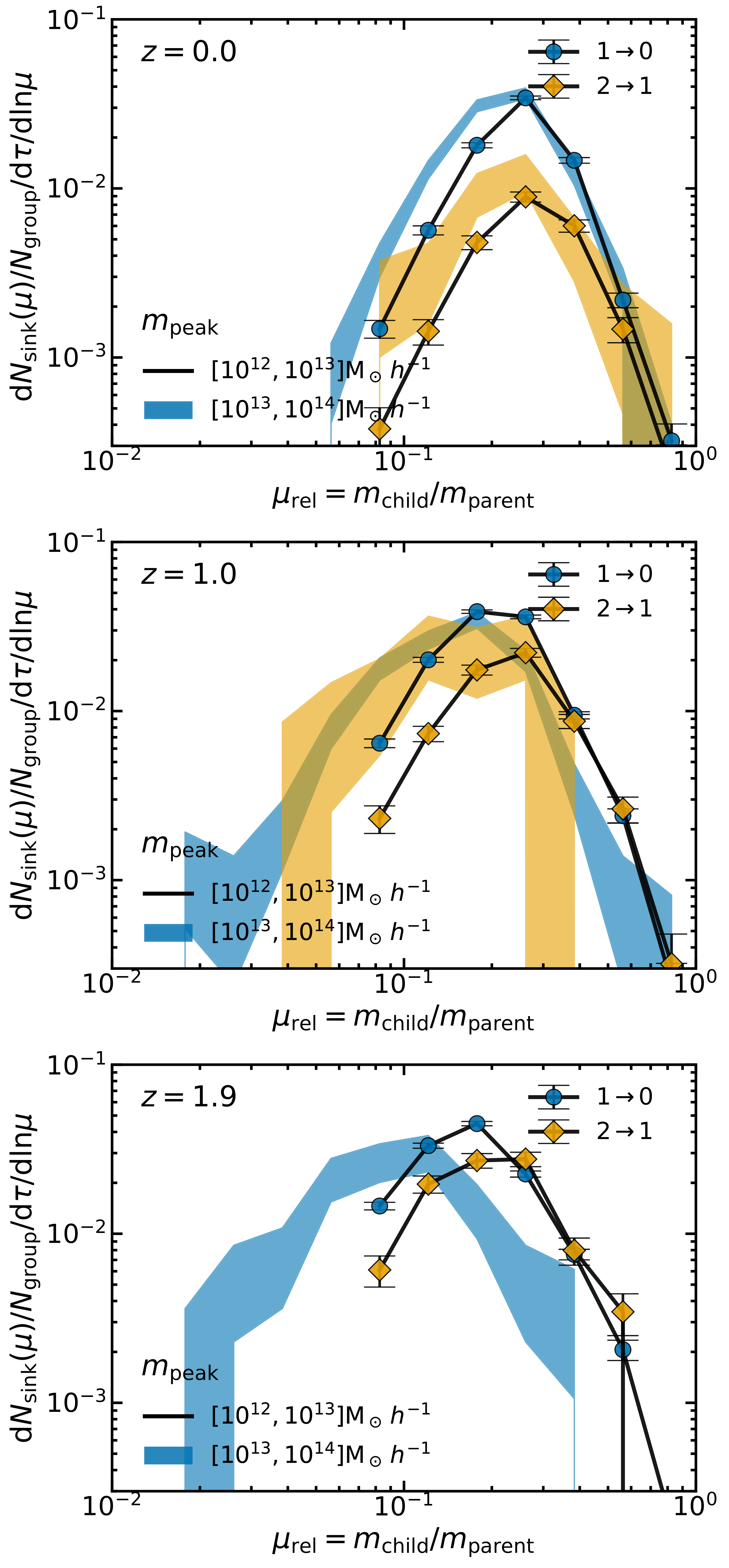}
    \caption{\textit{Top:} The sinking rate for subhalo groups at redshift 0. Lines with different colors and symbols indicate the sinking rate for subhalos of different levels. The shaded regions of different colors correspond to results from more massive groups, and their thickness indicates the uncertainty derived from the Poisson error on the mean. Here, the group mass is determined by the peak masses of the direct parents of these sunken subhalos. \textit{Middle:} The same as the top panel but for redshift 1. \textit{Bottom:} The same as the top panel but for redshift 1.9.}
    \label{fig:tot_sink_group}
\end{figure}

\subsection{Key Results and Interpretation}
\rev{Because the sinking rate can in principle vary over time, it is necessary that we explicitly specify the timescale over which we compute our sinking rate. In the following we present results measured over a timescale of $\sim 0.2 t_{\rm dyn}(z_{\rm 0})$ counted backward from the redshift of interest, $z_0$.}

Figure~\ref{fig:tot_sink_group} presents the group-level sinking rate $f_{\rm group}$ measured from two group mass bins at three different redshifts. The finite time window is chosen to be $\Delta t = 0.2 t_{\rm dyn}(z=0),0.24 t_{\rm dyn}(z=1),0.25 t_{\rm dyn}(z=1.9)$ respectively, with the exact number determined by the available snapshots around each time window. We find that the sinking events often occur between massive child subhalos ($\mu_{\rm rel}>0.1$) and their direct parents for $\ell=1,2$. This is expected because subhalos with large relative mass ratios typically experience stronger dynamical friction and their orbits decay more rapidly.
\begin{figure}[htbp]
    \centering
    \includegraphics[width=1.\linewidth]{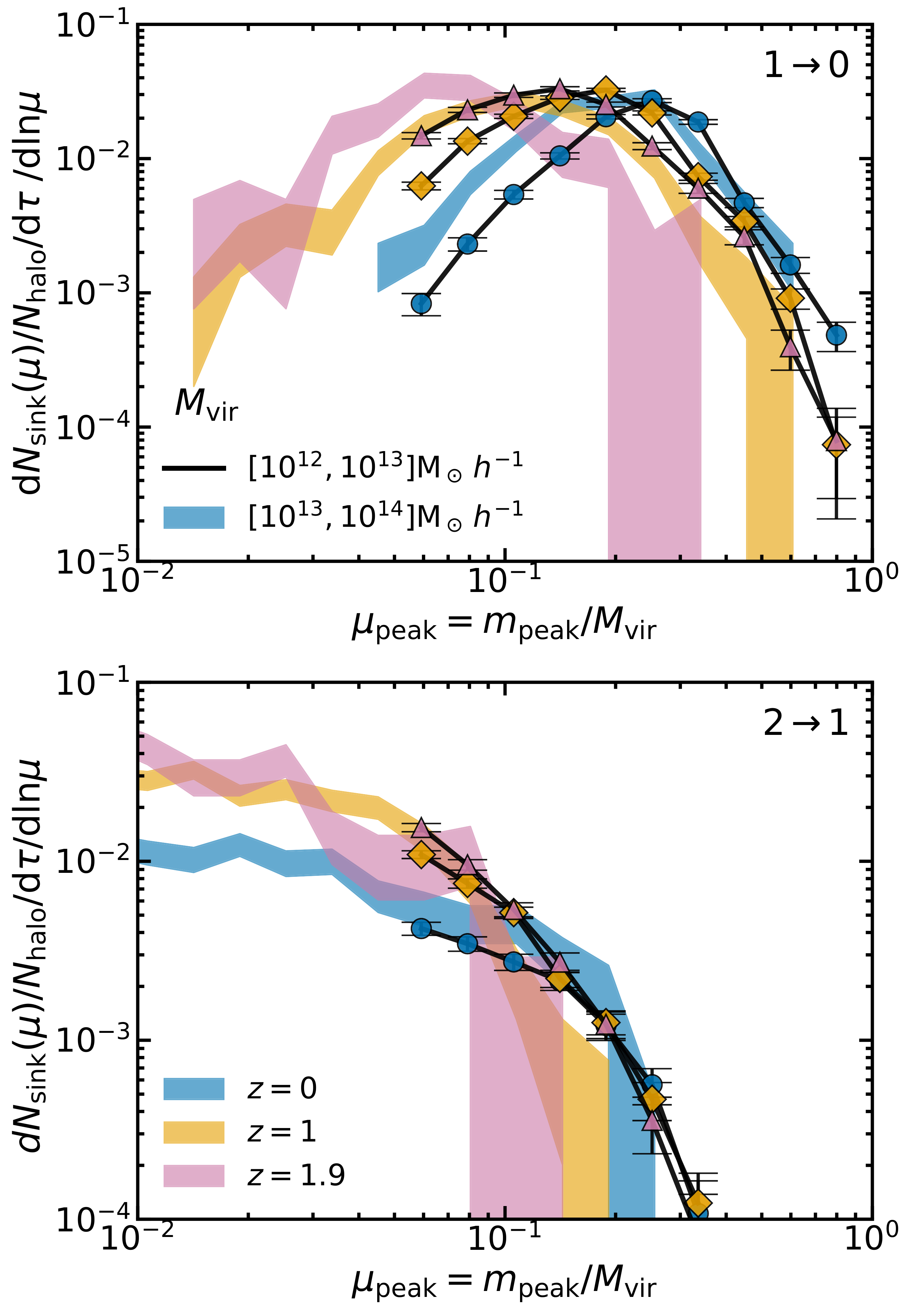}
    \caption{\textit{Top:} The sinking rate per halo for child subhalos at different redshifts. We used colored symbols to represent the sinking rate of level-1 subhalos at a given peak mass ratio at three different redshifts. The shaded regions in each plot indicate the sinking rate measured from more massive halo samples ($M_{\rm vir}$) with error bars. The error bars correspond to the Poisson noise derived from the total count of sunken subhalos in each peak mass ratio bin. \textit{Bottom:} Similar to the top panel, but measured from level-2 subhalos. }
    \label{fig:tot_sink_halo}
\end{figure}
Another key finding is that the amplitude of the level-2 sinking rate at $z=0$ is approximately a factor of four lower than that of level-1. However, the sinking rate of these two levels becomes comparable at high redshifts ($z=1$ and $z=1.9$). This suppression could be attributed to three possible factors:
\begin{itemize}
    \item Starvation of child subhalos. Once a satellite group first falls into a larger halo, it ceases to accrete new member subhalos. As a result, its member subhalo population may become largely exhausted by $z=0$, leading to a lower sinking rate.

    \item Tidal mass loss. Since deeper-level subhalos on average have larger mass ratios relative to their host halos at \rev{accretion} time~\cite{group_infall4}, and such high-mass-ratio subhalo populations are also shown to occupy more radial orbits~\cite{Li20}, they tend to experience stronger tidal mass-loss rates~\cite{artificial_disruption3}. As a result, deeper-level subhalos are more likely to become unresolved at the final redshift of interest, given the finite mass resolution of the simulation. Since we only count resolved sinking events in this work, the sinking rate of level‑2 subhalos may thus be partly hidden by resolution limitations compared with that of level-1 subhalos.

    \item Tidal heating. The tidal field of the host halo, as well as interactions with neighboring substructures, can heat the orbits of child subhalos within the subhalo group, delaying their orbital decay and effectively suppressing the sinking rate.
\end{itemize}

Figure~\ref{fig:tot_sink_halo} shows the halo-level sinking rate $f_{\rm halo}$ with the same choice of the finite time window as in Figure~\ref{fig:tot_sink_group}. The level-1 rate closely resembles $f_{\rm group}$, differing only in the mass definition of its group. However, the level-2 rate reveals a substantial population of low-mass-ratio sinking events (extending down to $\mu_{\rm peak} \sim 10^{-2}$). This behavior can be understood as a natural consequence of the underlying parent population. The halo-level sinking rate $f_{\rm halo,2}$ is essentially a convolution of two quantities: it combines the group-level sinking rate, $f_{\rm group,2}$, with the peak mass function of level-1 subhalos, $g_1(\mu_{\rm peak})$. Since the peak mass function steeply favors low-mass objects, the convolution is weighted toward small parent masses. Thus, even though $f_{\rm group,2}$ itself favors high $\mu_{\rm rel}$ and contributes little below $\mu_{\rm rel}\sim0.1$, the convolution with the abundant low-mass parent population drives $f_{\rm halo,2}$ to continue rising toward low $\mu_{\rm peak}$.
\begin{figure}[t]
    \centering
    \includegraphics[width=1.\linewidth]{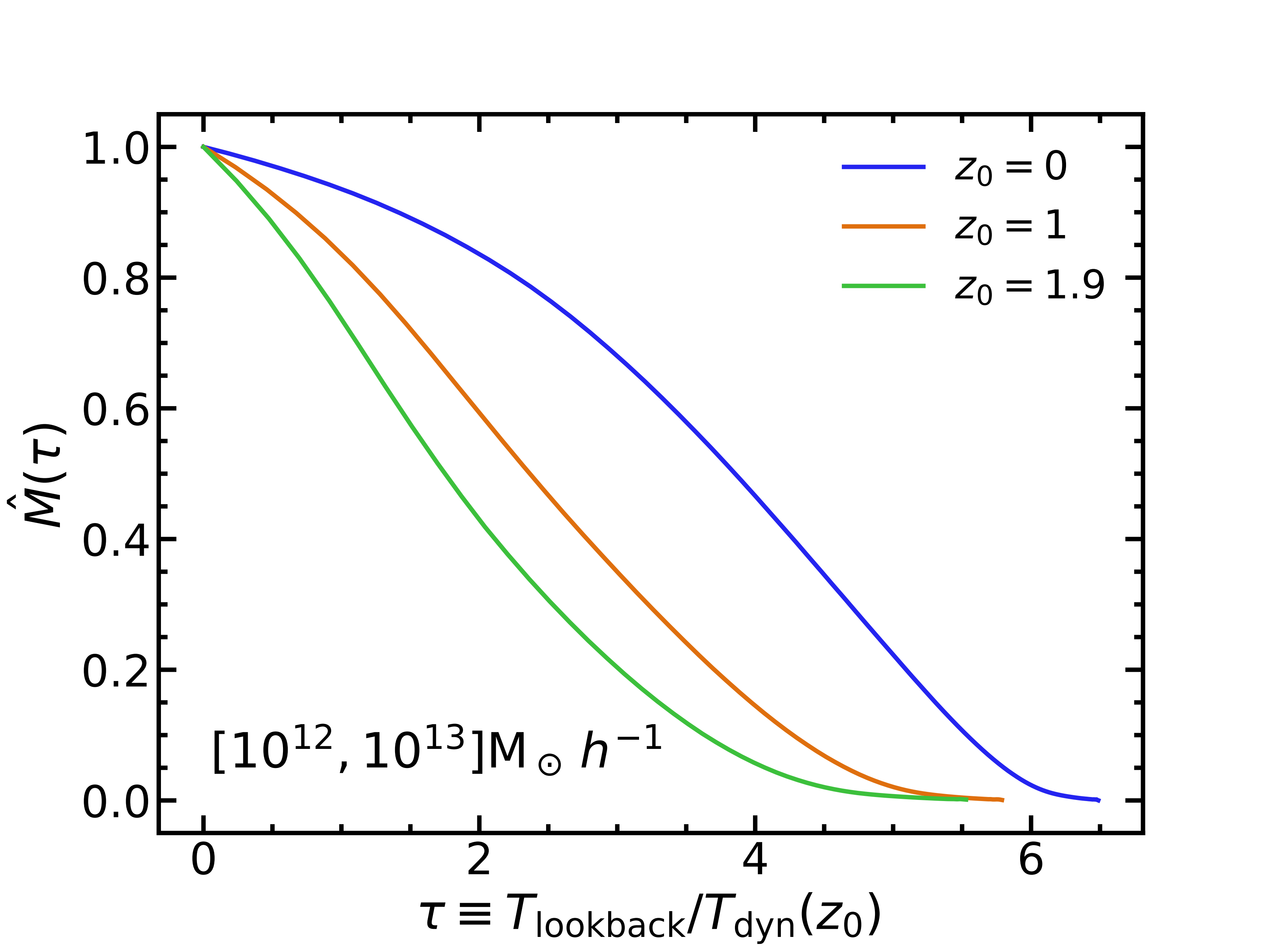}
    \caption{Normalized halo assembly history as a function of the lookback time. Colored lines illustrate the median mass assembly histories of halos in the mass range $[10^{12}, 10^{13}] \mathrm{M}_\odot\,h^{-1}$ across various redshifts. $\hat{M}(\tau)$ denotes the halo mass normalized by its final mass $M(z_0)$. Time is normalized to the dynamical timescale of the halos at their respective redshifts.}
    \label{fig:halo_growth}
\end{figure}

Furthermore, we find that the peak mass ratio distribution for level-1 sunken subhalos gradually shifts toward lower values with increasing redshift and host halo mass. This appears to contradict the self-similarity of the halo merger tree revealed by \cite{Fuyu,group_infall4}, which predicts a universal progenitor subhalo peak mass function across redshift. This apparent discrepancy can be understood from the different orbital phases of the subhalos since their accretion. Because lower mass ratio subhalos are on average accreted earlier~(see Figure 8 in \cite{group_infall4}), they also tend to sink earlier. As a result, lower mass ratio sinking events are more likely to happen in recently-formed halos, i.e., those that have assembled a large fraction of their masses (and subhalos) over a relatively short timescale. As shown in Figure~\ref{fig:halo_growth}, high redshift halos indeed assembled their masses more recently compared to lower-redshift halos, in accordance with the two-phase halo formation theory~\cite{Donghai}. Consequently, their sinking rate shifts systematically towards lower merger ratios. This trend is consistent with the results shown in Figure~\ref{fig:tot_sink_halo}. The same explanation applies to the host mass dependence, as high mass halos are formed more recently.

\begin{figure*}
    \centering
    \par
    \includegraphics[width=0.64\linewidth]{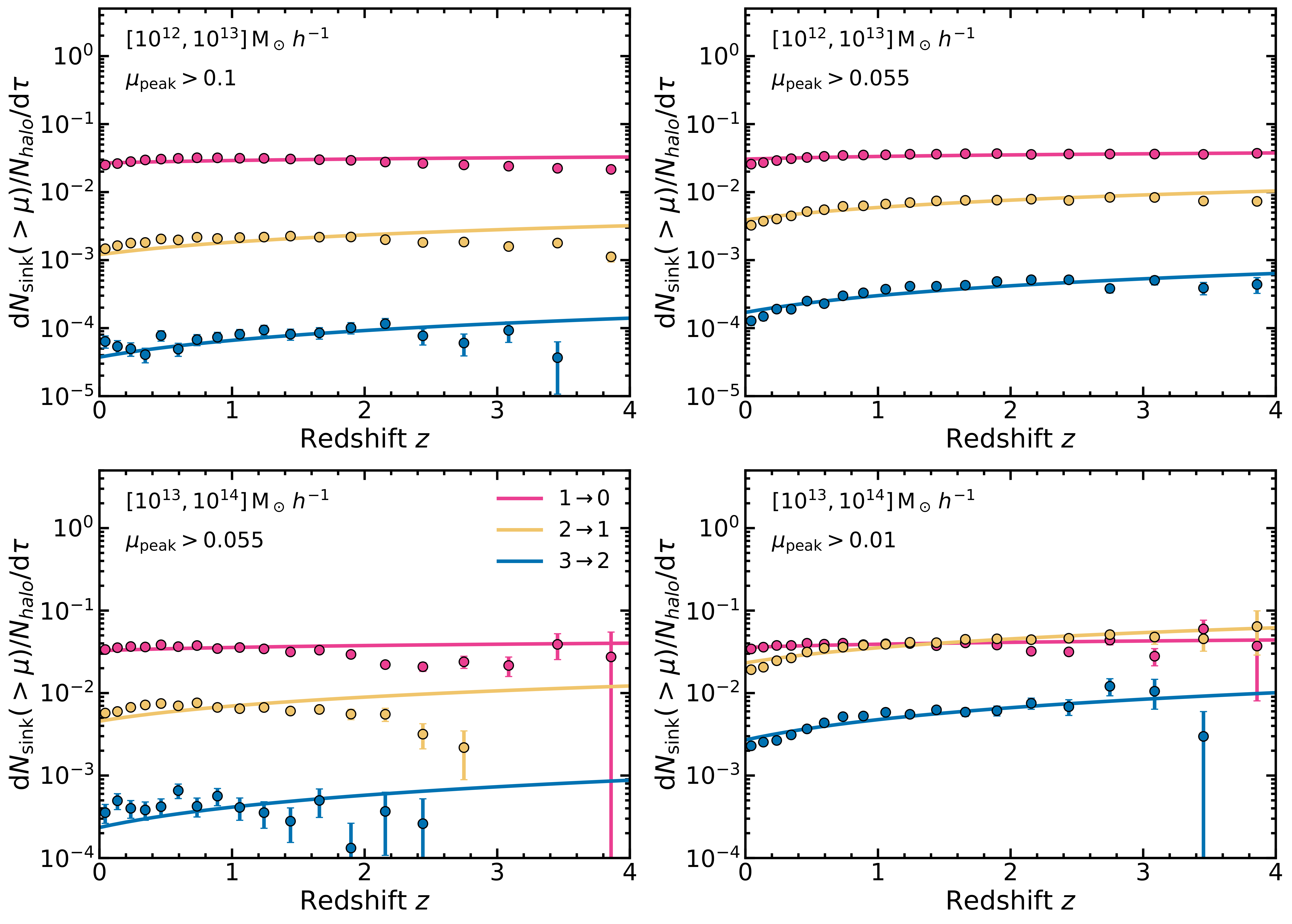}
    \caption{\textit{Upper panels:} The evolution of the cumulative halo-level sinking rate with respect to the redshift in host halos within $[10^{12},10^{13}]\mathrm{M}_\odot\,h^{-1}$ above two specific peak mass ratios. Colored solid lines in each panel show the best fit to the functional form $\propto (1+z)^{\alpha}$, with $\alpha$ being the fitted parameter. We use the Poisson error on the cumulative counts to calculate the error bars. \textit{Lower panels:} Similar to the upper panels, but for the sinking rate measured from more massive halos.}
    \label{fig:tot_sink_redshift}
\end{figure*}

Tracing the redshift evolution of $f_{\rm halo}$ above fixed mass-ratio thresholds (Figure~\ref{fig:tot_sink_redshift}), we find that $f_{\rm halo}$ increases steadily in the form of $ \propto (1+z)^{\alpha}$ for the first three levels. The slope $\alpha$ varies across the hierarchical levels with $\alpha=0.13,0.61,0.82$ for $\ell=1,2,3$ respectively. Notably, for level‑2 subhalos with $\mu_{\rm peak} > 10^{-2}$, their sinking rate becomes comparable to that of central--satellite (level-1) sinking events, highlighting the significant aggregate contribution of satellite--satellite mergers. It is expected that the $f_{\rm halo}$ with $\ell=3$ can also rival the central--satellite sinking rate when less massive subhalos are included in the analysis in simulations with higher particle resolution. The steady rise of the halo-level sinking rate for deep-level subhalos toward lower mass thresholds reveals that the satellite--satellite mergers become increasingly significant as we resolve subhalo \rev{populations} with progressively smaller peak mass ratios. 

\subsection{Comparison to Previous Work}

A. R. Wetzel et al.~\cite{merger_rate_Wetzel} report that satellite--satellite mergers constitute only a negligible fraction of the overall subhalo merger population compared with the central--satellite mergers. In contrast, we find a comparable satellite--satellite sinking rate in both Figures~\ref{fig:tot_sink_group} and~\ref{fig:tot_sink_redshift}. This apparent discrepancy, however, stems from differences in definitions and sample selection, rather than indicating a fundamental inconsistency.

A. R. Wetzel et al.~\cite{merger_rate_Wetzel} quantify the contribution of satellite--satellite mergers by measuring their fraction among all merger events whose descendant subhalos lie within a fixed bound mass bin (e.g., $[10^{12},10^{13}]\, \mathrm{M}_\odot\,h^{-1}$). They find that satellite--satellite mergers account for only $5\%$-$15\%$ of the total. This result is a natural consequence of their sample selection: within the descendant mass bins they consider, central subhalos dominate the population, while satellites constitute only a small fraction. Consequently, most merger events involving these descendants are central--satellite mergers, and satellite--satellite mergers are inevitably subdominant. While this statistic usefully characterizes the relative abundance of satellite--satellite mergers in a descendant-selected sample, it does not directly measure how frequently subhalos of shallower levels receive merged subhalos from deeper levels, nor does it reveal how the sinking rate varies across hierarchical levels.

Our group-level sinking rate adopts a similar descendant‑mass selection but takes a different normalization perspective. Rather than comparing counts to the total sunken population, we classify sinking events by the level $\ell$ of the sunken subhalo and normalize each class by the number of its parent groups, $N_{\rm group,\ell-1}$, i.e., the population that could host such an event. It directly quantifies the intrinsic sinking frequency for each hierarchical level within its own parent group and enables meaningful comparisons across different levels. Under this definition, the comparable group-level sinking rates of level-1 and level-2 subhalos at high redshift indicate that, for a given peak mass, a central subhalo is no more likely to undergo a sinking event than a level-1 satellite.

The halo-level sinking rate captures a fundamentally different statistic. It measures the total number of sinking events occurring within a host halo, aggregated across all hierarchical levels and normalized per halo. This definition is similar in spirit to the merger rate adopted by R. E. Angulo et al.~\cite{merger_rate_Angulo}. In this perspective, the steady increase of the level-2 sinking rate toward lower $\mu_{\rm peak}$—in contrast to the decreasing trend of level-1—reflects the growing contribution from the abundant population of low-mass level-1 subhalo groups.

Together, these two complementary definitions demonstrate that satellite--satellite mergers are not a negligible minority, but rather an important process that must be accounted for in any realistic model of subhalo evolution.


\section{The radial distribution of the sinking events}
\label{sec:spatial_distribution}

\begin{figure}[htbp]
    \centering
    \includegraphics[width=1.\linewidth]{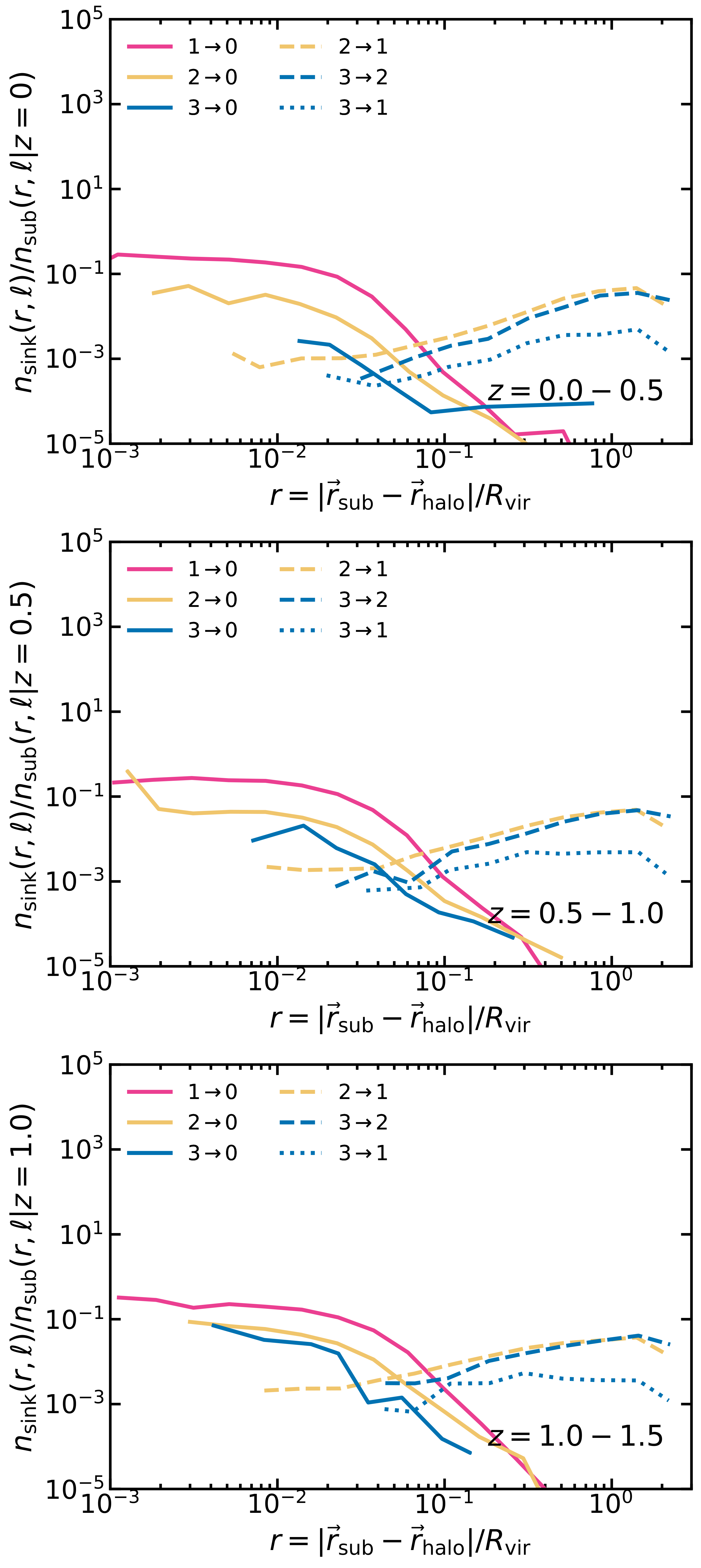}
    \caption{ Stacked radial probability function of sinking events for host halos of mass $[10^{12},10^{13}]\, \mathrm{M}_\odot\,h^{-1}$ at different redshifts. Radial distances are normalized by $R_{\rm vir}$ at the moment of sinking. Profiles are constructed from subhalos satisfying Equations~\ref{eq:sink_crit} and~\ref{eq:peak_crit}, and normalized by the number density profile. We categorize sinking events according to the hierarchical level of the sinking subhalo and the parent subhalo into which it sinks. The different categories are indicated by different line styles and colors.}
    \label{fig:spatial_sink_detail}
\end{figure}

Analyzing where sinking events occur provides crucial insights into where dynamical-friction-driven sinking is effective and where it is suppressed by multi-body interactions and tidal effects. Figure~\ref{fig:spatial_sink_detail} presents the stacked radial profiles of sinking events occurring in the host halos within the mass bin $[10^{12},10^{13}]\mathrm{M}_\odot\,h^{-1}$ across several redshift intervals. For each sinking event, the radial distance of the sunken subhalo is normalized by the virial radius of its host halo at the sinking moment, and the resulting stacked distribution is further normalized by the subhalo number density profile, which is measured from a specific snapshot using subhalos with peak mass greater than $100m_{\rm p}$. This normalization removes the baseline spatial distribution of the overall subhalo population, allowing us to isolate the preferred locations of sinking events. When classified by the nature of the sinking partner, the radial distributions of sinking events fall into two distinct categories.

Central--satellite mergers, in which a subhalo sinks directly into the central (level-0) subhalo, exhibit a centrally concentrated radial profile regardless of the levels of subhalos. Whether the sinking involves a level-1, level-2, or level-3 subhalo sinking into the central, the distribution peaks at $r\leq0.1R_{\rm vir}$ with an extended tail. The shape of this central concentration is remarkably consistent across levels, differing only in amplitude (see Figure~\ref{fig:spatial_sink_detail}). This universality reflects the common physical driver: dynamical friction operating deep within the host's potential well. Notably, a small fraction of these central--satellite mergers also occur at large radii ($r\gtrsim0.1R_{\rm vir}$). \rev{These distant events can be attributed to central subhalos with unusually diffuse spatial distributions. Their values of $\sigma_x$ lie in the top 5\% of the overall spatial dispersion distribution, confirming their anomalously diffuse nature.}

Satellite--satellite mergers, in which a subhalo sinks into a non-central parent (e.g., level-2$\rightarrow$level-1, level-3$\rightarrow$level-2, or level-3$\rightarrow$level-1), display a qualitatively different behavior. These events preferentially occur at intermediate to large radii, with broad peaks typically spanning $r/R_{\rm vir}\sim0.3$-$1.0$. This preference for the outer halo is robust across all redshifts examined.

\begin{figure}[htbp]
    \centering
    \includegraphics[width=1.\linewidth]{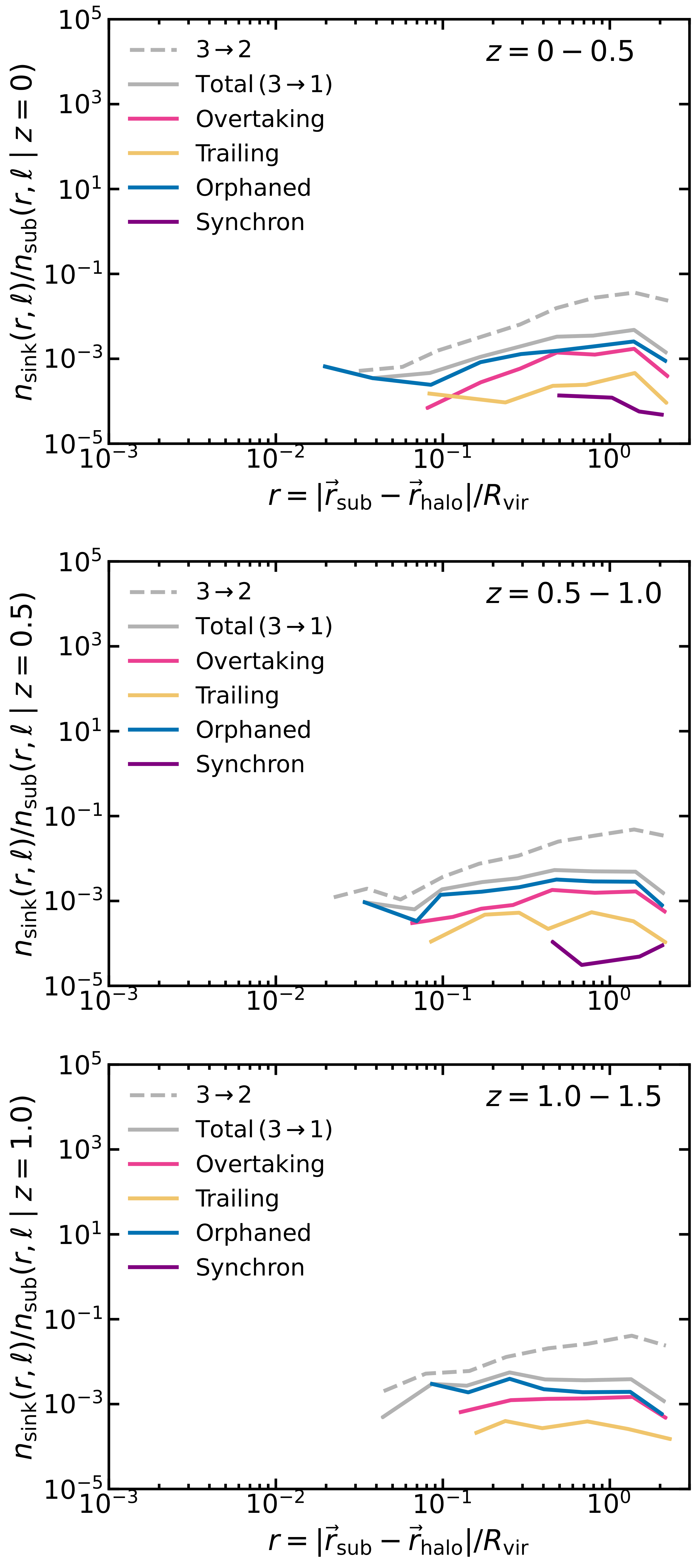}
    \caption{ Radial probability distribution of level-3$\rightarrow$level-1 sinking events, decomposed by cross--level sinking type. The four colored lines (excluding the gray ones) represent the four distinct cross--level pathways through which a level-3 subhalo sinks into a level-1 subhalo. The gray dashed line shows the overall distribution of all such events, while the gray solid line shows the distribution of next-level (level-3$\rightarrow$level-2) sinking events for comparison. Different panels correspond to different redshift intervals.
}
    \label{fig:sptial_sink_decomposition}
\end{figure}

The dominance of satellite--satellite mergers in the outer regions is consistent with previous findings \cite{merger_rate_Angulo, wetzel_clustering}. This suggests that subhalo groups near the center cannot retain their member subhalos long enough for orbital decay to complete.

Within this satellite--satellite category, we observe that the radial distributions of those sinking into their direct parents (i.e., level-2$\rightarrow$level-1 and level-3$\rightarrow$level-2) are remarkably consistent with each other, exhibiting nearly identical shapes across all epochs. In contrast, sinking events involving a level-3 subhalo sinking into a level-1 parent show a distinctly flatter radial profile at large radii. This flattening becomes more pronounced at higher redshifts. To understand this behavior, we decompose the level-3$\rightarrow$level-1 sinking events according to the cross--level classification introduced in Section~\ref{sec:sink_fate} (Figure~\ref{fig:sptial_sink_decomposition}). We find that the \textit{Orphaned Pathway} type dominates this channel across all radii, and its radial profile is significantly flatter than those of the other three types. Moreover, the shape of the \textit{Orphaned Pathway} distribution evolves with redshift, raising the question of whether this flattening trend is physical or numerical in origin. The issue can be resolved with higher-resolution simulations. In contrast, the second most abundant type, \textit{Overtaking Pathway}, exhibits a radial distribution whose shape still closely follows that of next-level (level-3$\rightarrow$level-2) sinking events.

Taken together, the radial probabilities highlight that the sinking events are intimately connected to the hierarchical nature of structure formation. While central--satellite mergers dominate the innermost halo regions, satellite--satellite mergers preferentially occupy the outskirts, underscoring the key roles played by \rev{group accretion}, tidal perturbations, and the multi-level subhalo hierarchy in governing where sinking events occur.

\rev{We emphasize that here we define the location of a sinking event as its onset location. Alternatively, one can choose to define the sinking event location at the center of the host subhalo. In this case, the spatial distribution of the central--satellite sinking profile will follow a delta-function by construction, and the satellite--satellite sinking will similarly be a deconvolved version of that presented above.}

\section{Resolution dependence of sinking detection and statistics}
\label{sec:robustness}

\rev{In this section we investigate how the identification and statistics of the sinking events depend on the numerical resolution of the simulation, in order to assess the robustness of the results.
We first examine the numerical scales that enter the sinking criterion, focusing on how the parent-core dispersions $(\sigma_x,\sigma_v)$ depend on subhalo bound mass and simulation resolution. Then we assess whether our current sinking rate statistics have converged across numerical resolutions. We further discuss the relation of our current sinking definition to some previous merger definitions, in light of the convergence study.}

\subsection{Resolution Dependence of Core Sizes}
\label{sec:discussion_resolution}
\begin{figure*}
    \centering
    \includegraphics[width=1.\linewidth]{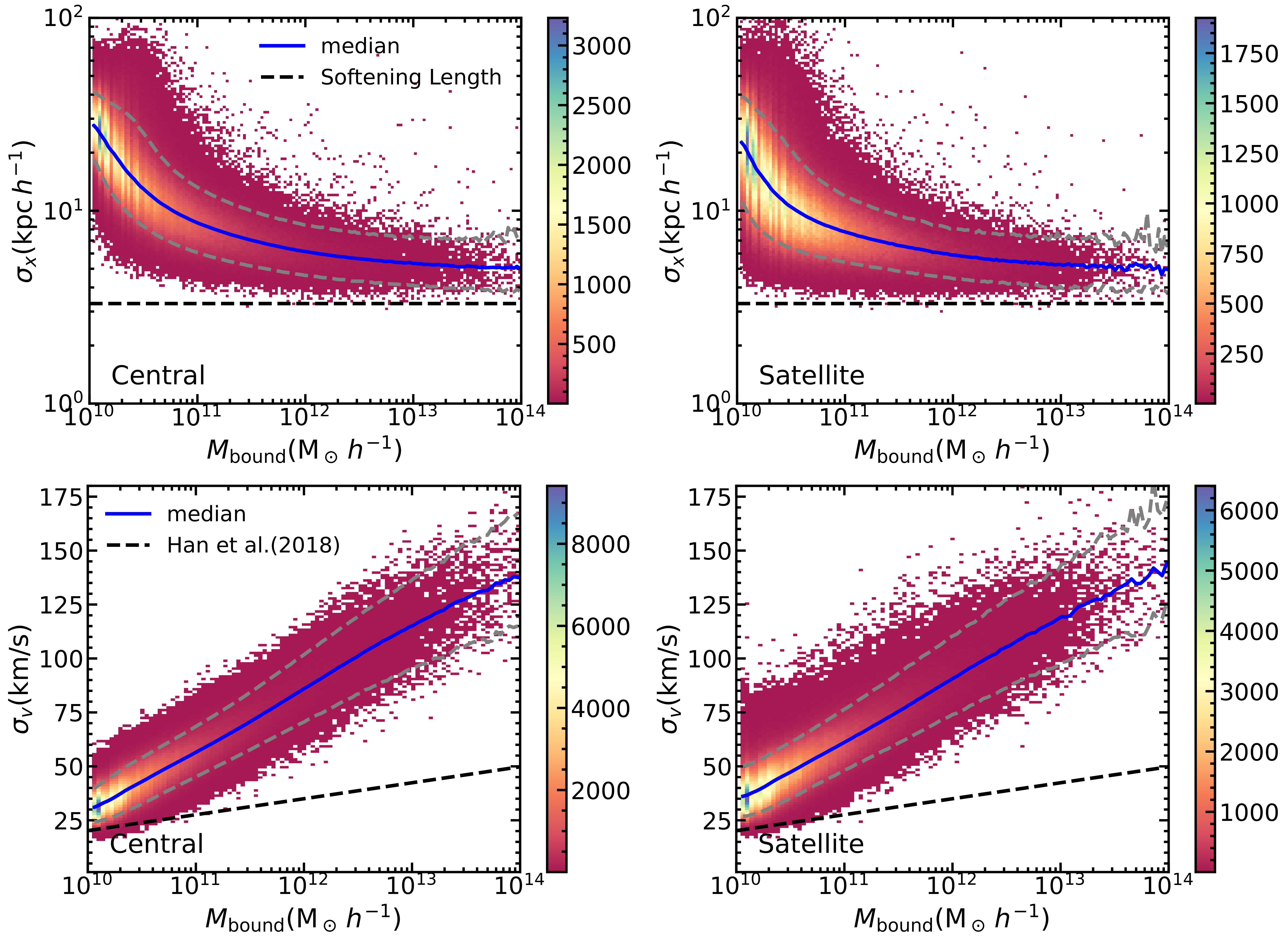}
    \caption{\textit{Upper panels:}  Spatial dispersion of the 20 most bound particles of subhalos at $z=0$ as a function of bound mass. The left and right panels correspond to central and satellite subhalos, respectively. The blue solid line indicates the median dispersion in each mass bin, and the black dashed line marks the simulation softening length. The region between the two gray dashed lines in each panel encloses the central 90\% (5th-95th percentile) of subhalos at a given mass. For clarity, only $10^5$ randomly chosen subhalo samples are shown, with colors indicating subhalo density. \textit{Lower panels:} Velocity dispersion of the 20 most bound particles of subhalos, following the same layout as the upper panels. The blue solid line represents the median of the overall velocity dispersion distribution, and the black dashed line shows the $\sigma_v-N_{\rm particle}$ fitting formula derived from the Millennium-II simulation in \cite{Han18}. As above, the gray dashed lines enclose the 5th-95th percentile range in each mass bin.}
    \label{fig:sigma_x_v_subhalo}
\end{figure*}

Figure~\ref{fig:sigma_x_v_subhalo} shows the distribution of $\sigma_x$ and $\sigma_v$ as a function of subhalo bound mass for both central and satellite subhalos. The median spatial dispersion $\sigma_x$ decreases with subhalo mass and asymptotes to a value above the force softening length for subhalos containing at least 200 particles. This implies that, when applying this method, the minimum requirement is that the parent subhalo must be well-resolved (e.g., $M_{\rm vir}\geq10^{12}\,\mathrm{M}_\odot\,h^{-1}$ or $m_{\rm parent}\geq10^{12}\,\mathrm{M}_\odot\,h^{-1}$ in this work) for the core scales to be reliably measured. The velocity dispersion $\sigma_v$, in contrast, increases with mass, reflecting the depth of the gravitational potential. Notably, the median values of both distributions are nearly identical for satellites and centrals, indicating that these core scales are \rev{not strongly affected by the host-halo tidal environment for the resolved subhalos considered here}.
These results establish that the stalling scales of orbital decay are set by the subhalo's intrinsic potential and the simulation's numerical resolution.

The inherent dependence of $\sigma_x$ and $\sigma_v$ on numerical resolution becomes evident when comparing our results with the higher-resolution Millennium-II analysis in J. Han et al.~\cite{Han18}, which adopts a softening length $\epsilon = 1\,\mathrm{kpc}\,h^{-1}$ and a particle mass $m_{\rm p}=6.89\times10^{6}\,\mathrm{M}_\odot\,h^{-1}$. In the $\sigma_x$ distribution, the median reported in J. Han et al.~\cite{Han18} asymptotes to approximately the softening length, whereas our median lies at about 1.5 times our softening length. This offset likely stems from differences in simulation codes, softening schemes, or the systematic shift induced by adopting different softening lengths. For $\sigma_v$, while the overall trends with subhalo mass are consistent with the fitting formula in J. Han et al.~\cite{Han18}, our measured velocity dispersions are systematically larger. This is a direct consequence of particle mass resolution: for a given subhalo mass, the 20 most bound particles (from which $\sigma_x$ and $\sigma_v$ are computed) sample a larger region of phase space than in a higher-resolution simulation, due to the larger particle mass in our simulation. This naturally leads to a higher velocity dispersion according to the rising velocity dispersion profile in the inner region of dark matter halos. 

The resolution dependence underscores that a sinking event identified at a given resolution \rev{does not occur at an identical physical scale at higher resolution. This introduces a resolution dependence into the spatial distribution of the sinking events. Despite this, the sinking rate statistics are largely unaffected, due to the robustness in identifying the sinking time, as we will show through a convergence study in the next subsection.} 


\subsection{Convergence of sinking rate statistics}
\label{sec:discussion_convergence}
\rev{In the previous subsection, we showed that the characteristic core scales
$(\sigma_x,\sigma_v)$ used by the sinking criterion depend on numerical
resolution. A higher-resolution simulation resolves a smaller and dynamically
colder parent core. For the same physical system, this can shift the snapshot
at which the condition $\delta_{\rm s}<2$ is first satisfied. At the same time,
improved mass resolution allows subhalos to remain resolved for longer, and
may therefore recover sinking events that would have been classified as orphans in
a lower resolution run. It is therefore necessary to examine whether the statistics of sinking events identified above remain robust at higher resolution.}

\rev{In order to assess the robustness of our results, we repeat the measurements of the sinking rate and spatial probability using a higher-resolution simulation. For this comparison, we use the M300 run from the Jiutian simulation suite~\cite{Jiutain_Overview}. This simulation evolves a periodic box of side length $300\,{\rm Mpc}\,h^{-1}$ with cosmological parameters $\Omega_{\rm M_0}=0.3111$ and $\sigma_8=0.8102$. The particle mass is $m_{\rm p}=1.005\times10^7\,{\rm M}_\odot\,h^{-1}$, approximately a factor of 55 lower than that of the L600 run, and the Plummer-equivalent gravitational softening length is $\epsilon=1\,{\rm kpc}\,h^{-1}$ in comoving units. We apply the same subhalo selection criteria as defined in Equations~\ref{eq:sink_crit} and~\ref{eq:peak_crit}. The finite time window used in measuring the sinking rate is kept identical to that adopted in the fiducial analysis.}
\begin{figure}[t]
    \centering
    \includegraphics[width=1.\linewidth]{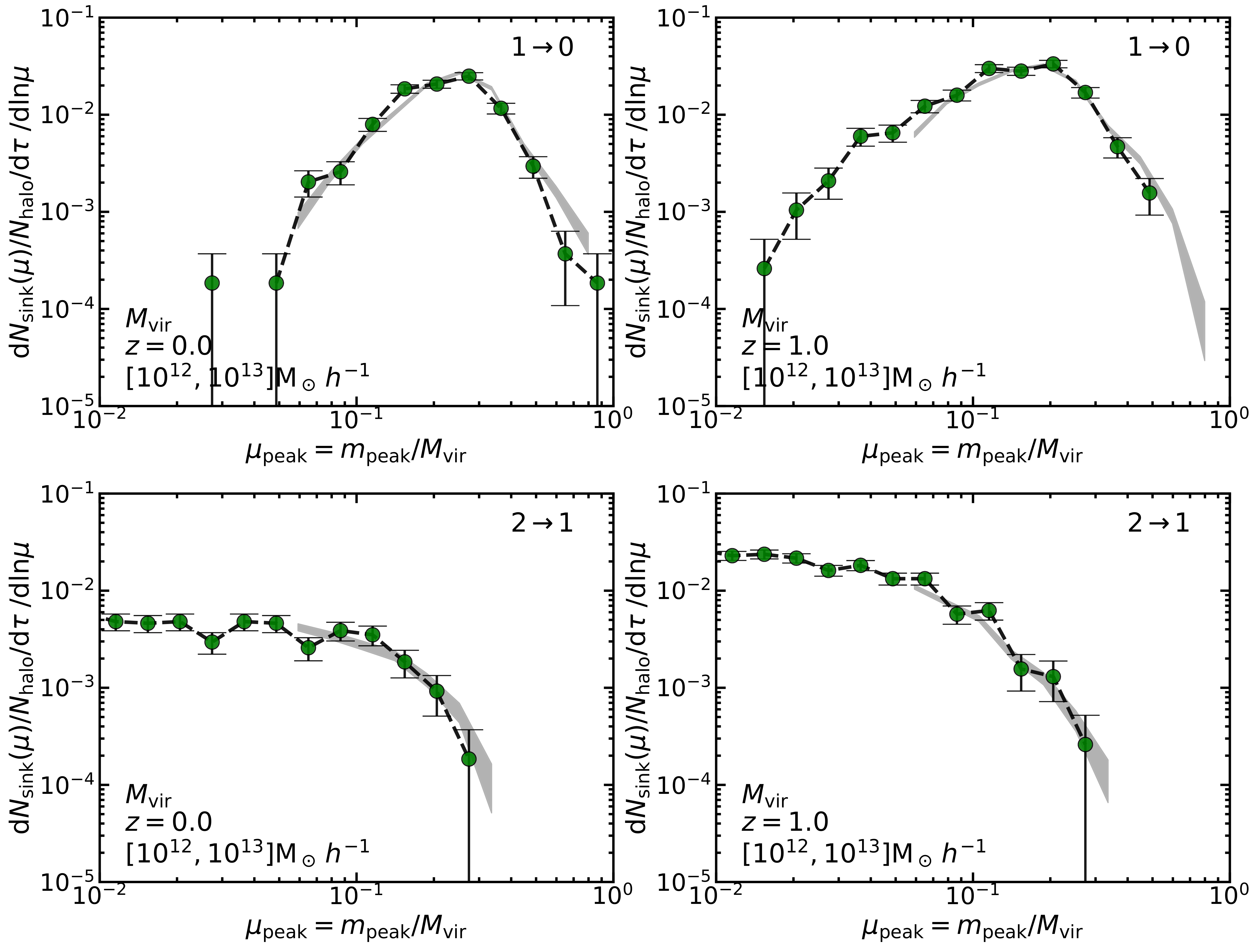}
    \caption{Halo-level sinking rates measured from L600 and Jiutian/M300 for host halos in the mass range $[10^{12},10^{13}]\,{\rm M}_\odot\,h^{-1}$. The upper panels show the results for level-1 subhalos at $z=0$ and $z=1$, while the lower panels show the corresponding results for level-2 subhalos. Green symbols with error bars show Jiutian/M300, and gray bands show L600.}
    \label{fig:M12_halolevel_compare}
\end{figure}

\rev{Figure~\ref{fig:M12_halolevel_compare} compares the halo-level sinking rate between the two simulations for host halos with $M_{\rm vir}=[10^{12},10^{13}]\,{\rm M}_\odot\,h^{-1}$, a regime that is most likely affected by numerical resolution effects in the original results. For both $\ell=1$ and $\ell=2$, the M300 measurements reproduce the main trends seen in L600. The level-1 sinking rate peaks at $\mu_{\rm peak}\gtrsim 0.1$ and decreases toward both lower and higher values of $\mu_{\rm peak}$, whereas the level-2 rate increases toward lower $\mu_{\rm peak}$. In the range of $\mu_{\rm peak}$ that is well populated in both simulations, the two datasets are generally consistent within the uncertainties. Quantitatively, the bin-by-bin differences are typically constrained to within $\sim20\%$, except for bins with large statistical uncertainties. A similar level of agreement (within $\sim20\%$) is also found for the more massive halo bin, $M_{\rm vir}=[10^{13},10^{14}]\,{\rm M}_\odot\,h^{-1}$.}

\begin{figure}[t]
    \centering
     \includegraphics[width=1.\linewidth]{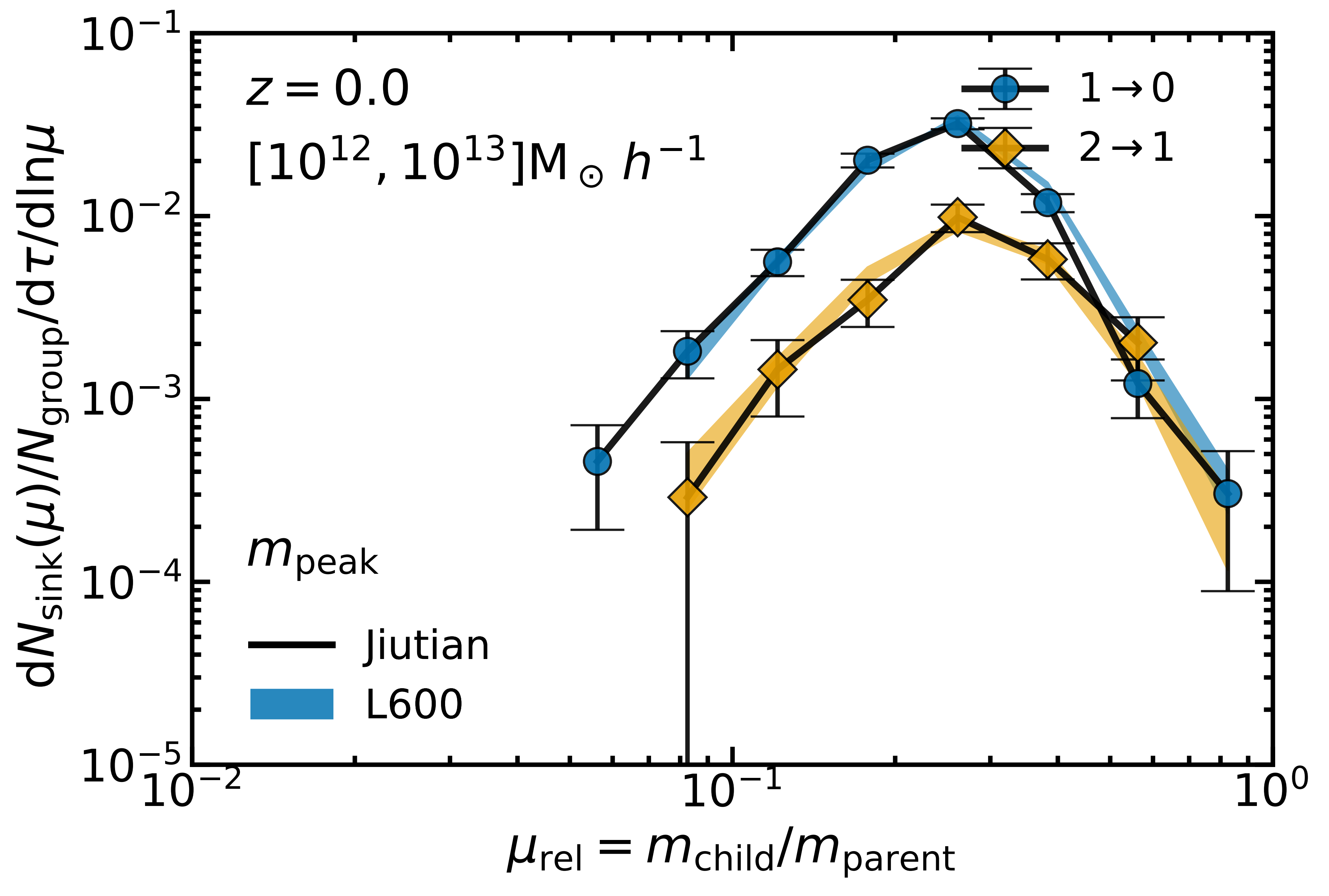}
    \caption{Group-level sinking rates at $z=0$ measured from L600 and Jiutian/M300 groups with masses in $[10^{12},10^{13}]\,{\rm M}_\odot\,h^{-1}$. Blue and orange symbols denote the level-1 and level-2 rates in Jiutian/M300, respectively; the shaded regions show the corresponding L600 measurements.}
    \label{fig:M12_grouplevel_compare}
\end{figure}

\rev{Figure~\ref{fig:M12_grouplevel_compare} also presents the corresponding group-level rates at $z=0$. The M300 results again follow the trends seen in L600, with better agreement between the two simulations and deviations typically within 15\%. The group-level rates for higher-mass groups are subject to larger statistical uncertainties but do not show evidence of significant differences.}

\begin{figure}[t]
    \centering
    \includegraphics[width=1.\linewidth]{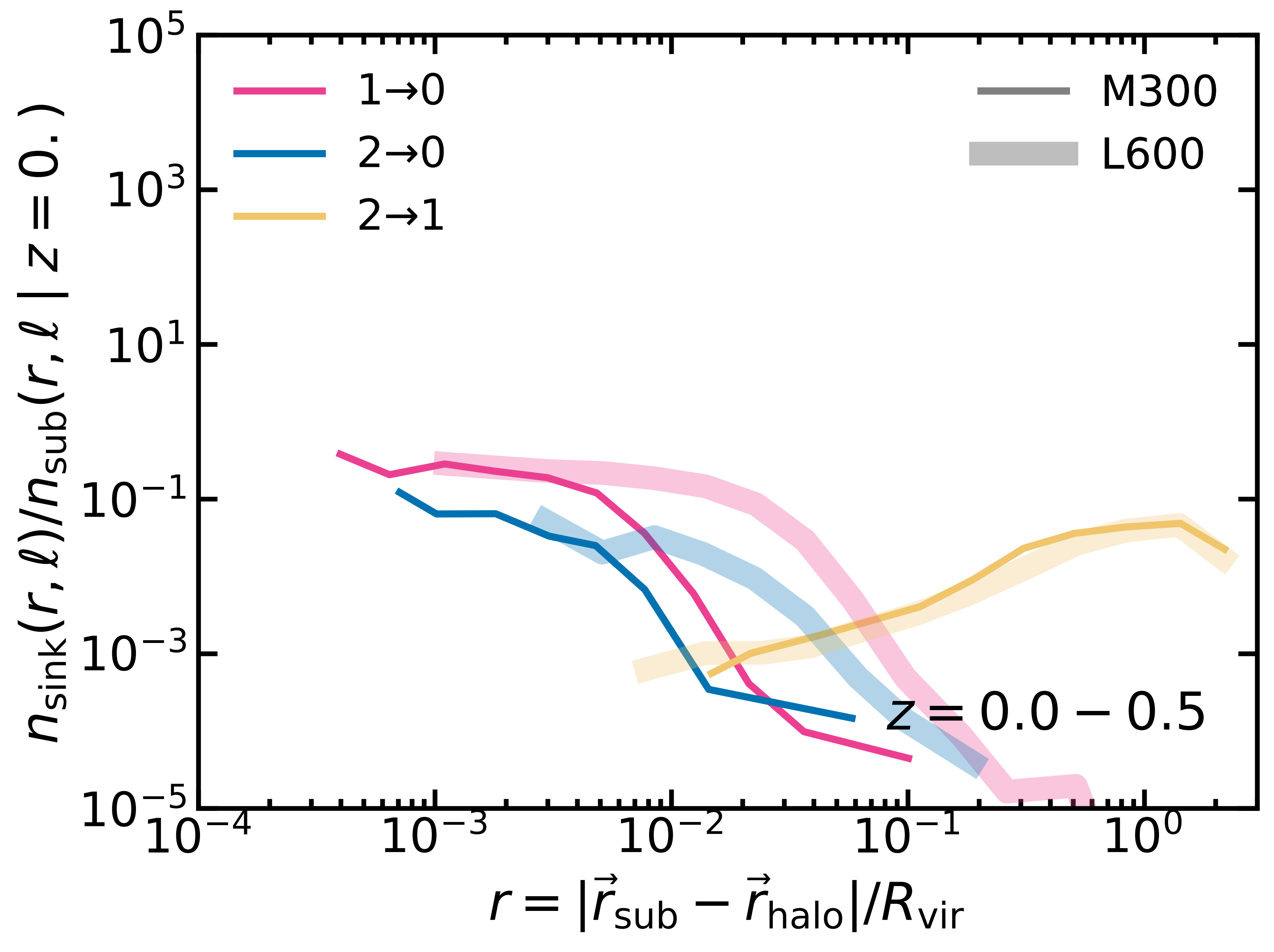}
    \caption{The radial probability distribution of sinking events in host halos with $M_{\rm vir}=[10^{12},10^{13}]\,{\rm M}_\odot\,h^{-1}$ for the L600(thick lines) and M300(thin lines) simulations. The panel shows the stacked radial distribution over the redshift interval $[0,0.5]$. Different colored lines indicate sinking events associated with different combinations of subhalo hierarchical levels.}
    \label{fig:Nsink_spatial}
\end{figure}

\rev{Figure~\ref{fig:Nsink_spatial} shows the radial probability distributions of sinking events stacked in the redshift range $0<z<0.5$. The bimodal structure remains clearly visible in the higher-resolution simulation. The satellite--satellite ($\ell = 2 \to \ell = 1$) component agrees well between the two simulations, with deviations typically within 25\%. This indicates that the radial distribution of satellite--satellite mergers is largely insensitive to numerical resolution, as it is expected to be primarily determined by the radial distribution of the parent subhalos and the tidal influence of the host halo.}

\rev{A notable difference is that the central--satellite distribution systematically shifts toward the inner regions of dark matter halos, with essentially all events occurring within $r < 0.1\,R_{\rm vir}$. This shift reflects the smaller (more compact) central subhalo core in the higher-resolution simulation compared to the lower-resolution case, consistent with our previous analysis of resolution-dependent core properties. Consequently, the radial distribution becomes better resolved in the higher resolution run, while the lower resolution run can be interpreted as a smeared version of the higher resolution one. We further emphasize that this resolution dependence is only introduced when we study the distribution of the onset of sinking events. When the sinking events are defined at the center of the host subhalo, the central--satellite distribution becomes a delta function by construction, irrespective of resolution.}

\rev{Despite the differences in the radial distribution of central--satellite mergers, the population-level sinking rate for level-1 subhalos remains well converged between simulations. This can be attributed to the fact that the transition from the decaying to the stalling phase occurs rapidly, on a timescale much shorter than the halo dynamical time as shown in Figure~\ref{fig:ds_evolution}. As a result, the integrated sinking rate is relatively insensitive to the precise spatial location of individual events, even though their detailed radial distribution varies with resolution. This implies that, provided the simulation can resolve the relevant phase transition at the mass scale of interest, robust measurements of sinking statistics can still be obtained.}

\rev{A complication in understanding the convergence of the sinking rate arises from the orphan subhalos. It may be expected that as the resolution increases, some orphan subhalos will be identified as surviving ones, and thus could boost the sinking rate. The fact that we do not see a significant boost in our convergence test can be understood from two aspects. First of all, according to the universal subhalo distribution model, about half of the accreted subhalos at a given peak mass can be identified as effectively disrupted irrespective of simulation resolution~\cite{Han16}, corresponding to a population of extreme mass loss rates~\cite{artificial_disruption3}. Varying the resolution thus does not change the abundance of this population, which constitutes the majority of the orphan population at a given peak mass. For the surviving population, the majority ($>$90\%) only experiences moderate (typically a factor of $\sim 10$) mass loss (see Figure 7 in \cite{Han16}). Moreover, the sinking rate is dominated by major mergers with mass ratios above $\sim 0.1$ (see Figure~\ref{fig:M12_grouplevel_compare}). Combining the two factors, a rough estimate shows that resolving a typical sunken subhalo with $20$ particles requires its direct host halo to have at least 2000 particles, which corresponds to $10^{12}\msunh$ for our fiducial L600 simulation. As a result, it is not surprising that our results show good convergence in the group-level sinking rate even for the $10^{12}-10^{13}\msunh$ bin, let alone the $10^{13}-10^{14}\msunh$ bin resolved with many more particles. 
}


\subsection{Comparison with Conventional Merger Criteria}
\begin{figure*}
    \centering
    \includegraphics[width=1.\textwidth]{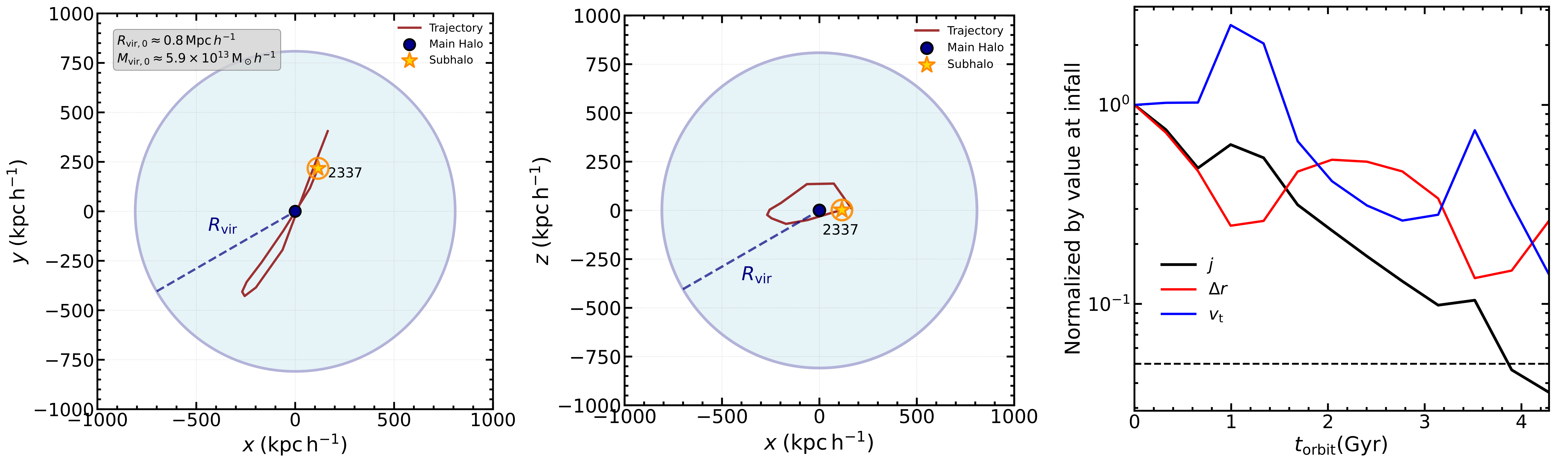}
    \caption{\textit{Left:} $xy$-plane projection of the orbital trajectory of a level-1 subhalo. The red solid line traces the trajectory of this level-1 subhalo (directly orbiting the main halo) from its first crossing of the host's virial radius to the present epoch. The main halo center at $z=0$ is marked by a black dot, and its virial radius ($R_{\rm vir,0} \approx 0.8$ Mpc $h^{-1}$, corresponding to a mass $M_{\rm vir,0} \approx 5.9 \times 10^{13} \mathrm{M}_\odot\,h^{-1}$) is shown as a blue circle. The subhalo's current position (at $z=0$) is indicated by a gold star. We use the orange-solid circles to show the half-mass radius of this subhalo at the final spatial location during the evolution. Meanwhile, we also display the number of bound particles at that time near the orange circles. At present, the subhalo resides at a normalized separation $\Delta r / R_{\rm vir} = 0.3$ from the halo center, while its orbital angular momentum has decayed to only $\sim 3.4\%$ of the value it possessed when it initially crossed the virial boundary. \textit{Middle}: $xz$-plane projection of the same subhalo. \textit{Right}: Evolution of several orbital parameters of the level-1 subhalo, normalized by their values at the time of \rev{accretion}. The black solid line shows the evolution of the specific angular momentum, while the red and blue lines denote the radial distance and tangential velocity, respectively, computed from the position and velocity averaged over bound particles. The intrinsic spin of the subhalo contributes little to the specific angular momentum during evolution.}
    \label{fig:case_study_J}
\end{figure*}

Although the sinking time is not sensitive to the spatial resolution of the simulation, different approaches to defining a merger can still produce systematically different answers. In particular, the results in Section~\ref{sec:discussion_resolution} highlight that the endpoint of orbital decay is not a universal fraction of the host virial radius $R_{\rm vir}$. Instead, the ratio $\sigma_x/R_{\rm vir}$ varies systematically across mass and resolution. Therefore, the sinking events identified by \hbt do not always correspond to those identified by arbitrarily proposed static thresholds, such as a subhalo crossing $0.1R_{\rm vir}$ or losing a fixed fraction (e.g., 95\%) of its initial specific angular momentum.  Figure~\ref{fig:case_study_J} illustrates this clearly: a subhalo in our simulation had lost 96\% of its initial specific angular momentum yet remained at $0.3R_{\rm vir}$ from the host center. A fixed physical threshold, therefore, unavoidably over-merges systems of different masses and resolutions, which explains why orbital evolution often continues well beyond the merger times identified by such rules, as also noted by J. M. Solanes et al.~\cite{merger_Solanes}.

\section{Discussion} 
\label{sec:disucssion}

\subsection{Reverse-Link Candidate Events}
\label{sec:reverse-link}
\rev{The preceding analysis uses the original single-sided {\hbt} sinking criterion. A recent extension to the \hbt algorithm, \hbtherons, modifies the sinking algorithm to account for cases in which a host subhalo may sink into the core of its satellite, so that the two still appear as a single object in phase space. Here we examine the events that would be added by such a reverse sinking criterion, in order to determine whether they trace the same orbital-decay process as the main sample.}
\subsubsection{A Representative Reverse Sinking Event}
\label{sec:reverse_case}

\rev{
To isolate the additional events introduced by the bidirectional criterion used in \hbtherons \cite{Victor}, we compare the original parent-normalized phase-space distance with the corresponding child-normalized distance:}

\begin{align}
\delta_{\rm s,p} &=
\frac{\lvert \mathbf{x}_{\rm c} - \mathbf{x}_{\rm p} \rvert}{\sigma_{x,\rm p}}
+ \frac{\lvert \mathbf{v}_{\rm c} - \mathbf{v}_{\rm p} \rvert}{\sigma_{v,\rm p}} ,
\\
\delta_{\rm s,c} &=
\frac{\lvert \mathbf{x}_{\rm p} - \mathbf{x}_{\rm c} \rvert}{\sigma_{x,\rm c}}
+ \frac{\lvert \mathbf{v}_{\rm p} - \mathbf{v}_{\rm c} \rvert}{\sigma_{v,\rm c}}.
\end{align}
\rev{where $\delta_{\rm s,p}$ is identical to the original $\delta_{\rm s}$ in Equation~\ref{eq:orig_sink_algorithm}, while $\delta_{\rm s,c}$ measures the distance from the parent to the child's core. This bidirectional approach expands the set of candidate events identified. Critically, only those subhalos that exclusively satisfy the reverse-link criterion ($\delta_{\rm s,c}<2$) while not simultaneously satisfying the original criterion ($\delta_{\rm s,p}<2$) during evolution are considered the additional sample introduced by this adaptation.}

\rev{To capture and accurately track this distinct sample of exclusively reverse-linked subhalos, a dedicated {\hbt} run is performed on the L600 simulation with the \texttt{MergeTrappedParticle} option disabled. This specific configuration 
allows the continuous tracking of both $\delta_{\rm s,p}$ and $\delta_{\rm s,c}$ for a subhalo even after an initial sinking criterion is met, until it eventually becomes an orphan subhalo. This capability enables the unambiguous identification of systems that only cross the $\delta_{\rm s,c}<2$ threshold without ever satisfying $\delta_{\rm s,p}<2$ during their observable evolution, thus precisely isolating the sample introduced by the reverse-link criterion.}

\begin{figure}[t]
    \centering
    \includegraphics[width=0.45\textwidth]{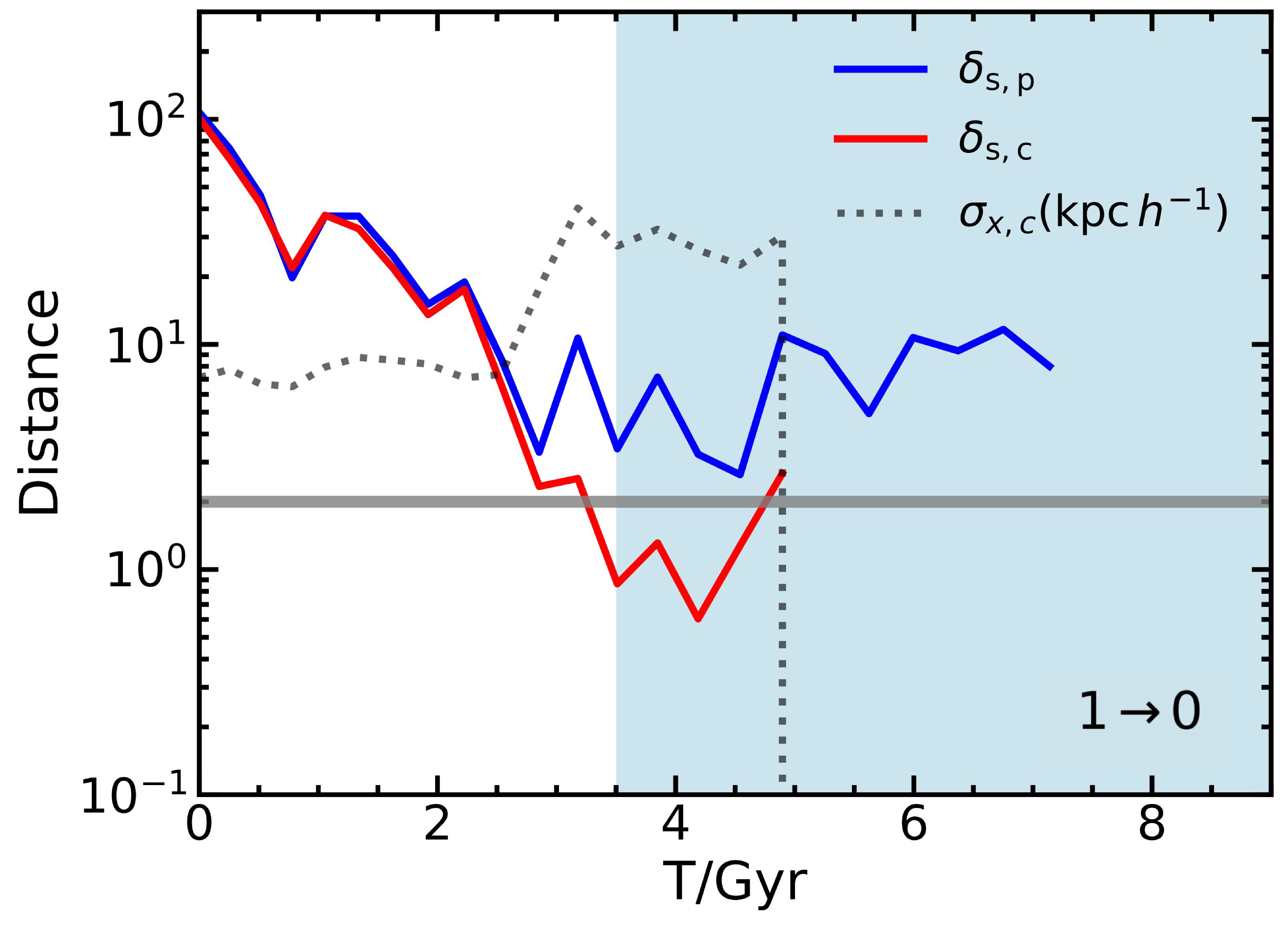}
    \caption{Evolution of $\delta_{\rm s,p}$ and $\delta_{\rm s,c}$ for a subhalo identified only by the reverse-link criterion (from the dedicated {\hbt} run in Section~\ref{sec:reverse-link}). The red and blue solid lines trace the evolution of $\delta_{\rm s,c}$ and $\delta_{\rm s,p}$, respectively. The gray dotted line shows the evolution of the spatial dispersion ($\sigma_{x,\rm c}$) of the child subhalo's core, while the gray solid line marks the threshold value of 2. The blue shaded region is used to represent the evolution after the moment satisfying the reverse-link criterion. The time axis begins at the snapshot when both the child subhalo and its direct parent are first recorded. The termination of the red line indicates that the child subhalo subsequently becomes unresolved. }
    \label{fig:reverse_study}
\end{figure}

\rev{Figure~\ref{fig:reverse_study} further displays the evolution of a representative level-1 subhalo that is identified by the reverse‑link criterion. This subhalo also experiences a rapid inspiral phase lasting several orbital periods. Its $\delta_{\rm s,p}$, however, stalls at a value slightly above the threshold of 2. Meanwhile, near this point, the internal spatial dispersion of the subhalo’s own core grows substantially. This growth is driven by the tidal heating when approaching the center of the host halo and in turn continuously reduces $\delta_{\rm s,c}$. This example illustrates the physical origin of reverse-link events. While the subhalo initially undergoes orbital decay, the subsequent decrease in $\delta_{\rm s,c}$ is primarily driven by the expansion of the core of child subhalos caused by tidal heating rather than by continued inspiral.} \rev{This case therefore motivates the statistical test below: we need to determine whether reverse-link events are generally associated with tidal heating, or whether they represent the same phase-space mixing process as the original sinking sample.}

\subsubsection{Driving Mechanisms: Orbital Decay versus Tidal Heating}
\label{sec:discussion_mechanism}

\rev{Figure~\ref{fig:reverse_study} presents a representative reverse-link event, suggesting that tidal heating may play an important role in driving $\delta_{\rm s,c}$ below the sinking threshold. We therefore investigate whether this behavior is common among reverse-link events and whether it also characterizes the sunken sample identified by the original {\hbt} criterion. To this end, we perform a statistical comparison of the structural evolution of subhalos across different samples of candidates.}

\rev{We divide all bidirectional-criterion candidates into three samples. The child-dispersion-driven sample contains events for which only the reverse-link criterion ($\delta_{\rm s,c}<2$) is satisfied. The parent-dispersion-driven sample contains events identified only through the original criterion ($\delta_{\rm s,p}<2$). The doubly identified sample contains events that satisfy both criteria during their evolution.}

\begin{figure*}
\centering
\includegraphics[width=0.8\linewidth]{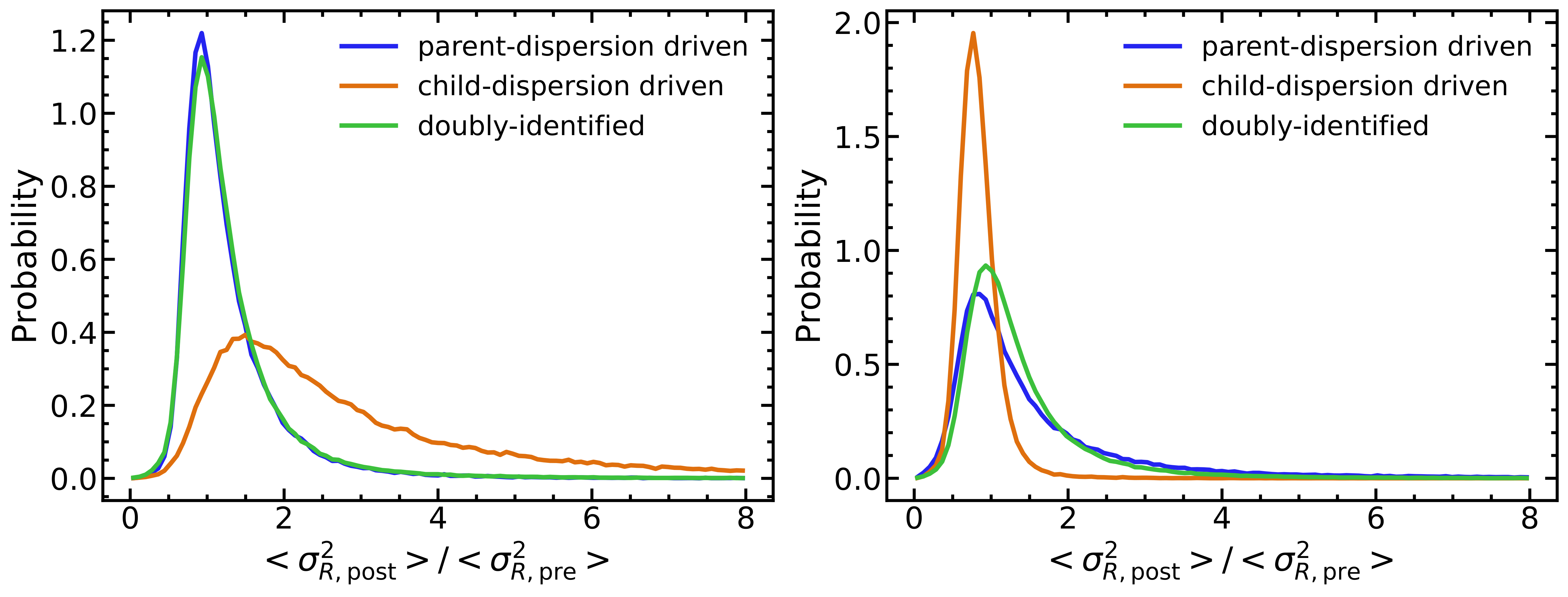}
\vspace{-1em}
\caption{Distribution of the ratio of the mean squared spatial dispersions measured after and before the criterion is satisfied. Colored solid lines distinguish \rev{the three samples defined in the text}. The sample includes all the candidates from halos with $M_{\rm vir}\geq10^{12}\mathrm{M}_\odot\,h^{-1}$, regardless of redshift.}
\label{fig:amplified_factor}
\end{figure*}

\rev{
To quantify the importance of tidal heating, we measure the change in the
spatial dispersion of subhalo cores across the relevant criterion-crossing
snapshot. We characterize this change by the ratio
$\langle\sigma^2_{R,\rm post}\rangle /
\langle\sigma^2_{R,\rm pre}\rangle$, where
$\langle\sigma^2_{R,\rm pre}\rangle$ and
$\langle\sigma^2_{R,\rm post}\rangle$ denote the average squared spatial
dispersion measured over the three snapshots immediately before and after this
criterion-crossing snapshot, respectively. The resulting distributions for the
three samples are shown in Figure~\ref{fig:amplified_factor}.
}

In child‑dispersion-driven cases, the majority ($>70\%$) of candidates experience significant tidal heating, with their $\sigma^2_R$ values increasing by a factor of $\sim2$ after satisfying the reverse-link criterion, while their parent subhalos remain nearly intact and compact. This confirms that reverse-link events are primarily driven by heating at the interface between the two evolutionary phases.

Conversely, subhalos in the parent-dispersion-driven and doubly identified samples remain compact throughout. Although the spatial dispersion of their parent subhalos may expand slightly, the child subhalos themselves show no evidence of tidal heating. In these cases, the transition between the two‑stage evolution is driven primarily by orbital decay rather than by heating.

\begin{figure}
\centering
\includegraphics[width=0.9\linewidth]{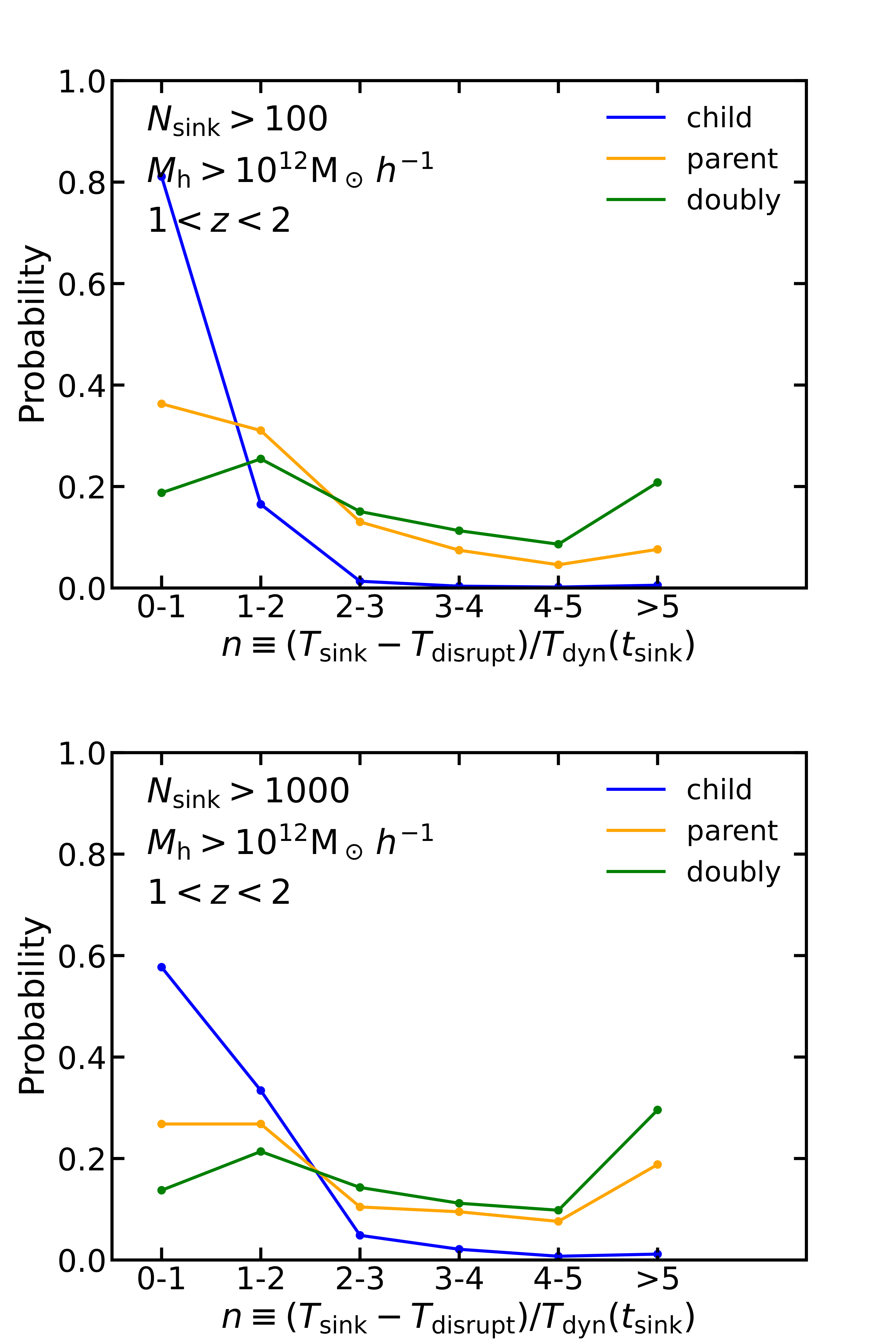}
\caption{\rev{\textit{Upper panel:} The survival timescale for subhalos that satisfy the criterion between $1<z<2$, considering only those with at least 100 bound particles at the snapshot crossing the threshold. The survival time is normalized by the dynamical timescale of the dark matter halo at the sinking time. Solid lines, differentiated by color in each plot, represent results measured from three distinct subhalo samples. \textit{Lower panel:} The corresponding results for the more restrictive sample of candidates with at least 1000 bound particles.}}
\label{fig:survival_time}
\end{figure}

\rev{The difference in the underlying physical mechanisms is further reflected in the survival times of the three samples, shown in Figure~\ref{fig:survival_time}. The survival time is defined as the time between sinking and disruption. The majority of child-dispersion-driven subhalos become unresolved within approximately 0--2 dynamical times after they satisfy the reverse-link criterion. In contrast, the parent-dispersion-driven and doubly identified samples exhibit substantially longer survival times, with extended tails reaching beyond five dynamical times.}

\rev{These results suggest that the reverse-link candidate events largely correspond to subhalos undergoing tidal heating that become unresolved shortly after satisfying the reverse-link criterion. Therefore, excluding these candidates has only transient impacts on the statistics of the resulting subhalo catalog. Our results also warn against some recent attempts that propose to account for subhalo sinking by identifying the heating event of the satellite subhalo~(e.g., \cite{Bloodhound}). Such candidates should be treated separately from the sinking events defined by the original {\hbt} criterion, and they contribute only a minor and largely transient population.} 


\subsection{Implications for Galaxy Evolution}
\rev{The statistical properties of the sinking rate have potential applications across several areas of galaxy formation and evolution, providing a statistical description of the long-term dynamical evolution of satellite galaxies within their hosts.}
\rev{
The dominance of next-level sinking pathways suggests that galaxy mergers are not always driven by the direct mergers between satellites and the central galaxy. Instead, hierarchical evolution and satellite--satellite interactions may also make a substantial contribution, and attributing all mergers to the central would significantly overestimate the growth rate of the central galaxy. As the morphological evolution and black hole growth rate are assumed to be determined by the merger rate of galaxies~(e.g., \cite{BH_merger1,BH_merger2,Morphology_merger1,Tan2026}), properly accounting for the different merger channels requires new calibration of the galaxy and black hole growth recipes. }

\rev{The hierarchical sinking statistics may also provide insight into merger-related AGN activity in dense environments. Observational studies have shown that AGN activity in groups and clusters depends on both redshift and cluster-centric radius, with high-redshift systems and cluster outskirts exhibiting enhanced AGN incidence or stronger AGN power, particularly among IR-selected AGN \cite{AGN_Trigger1,AGN_Trigger2,AGN_Trigger3}. Since IR-selected AGN efficiently trace dust-obscured accretion, a phase often linked to gas-rich mergers and merger-driven inflows~\cite{IR_AGN1,IR_AGN2,IR_AGN3}, they provide a natural population in which merger-induced gas inflows may play an important role. In this context, our finding that satellite--satellite sinking preferentially occurs in the outer regions of host halos and becomes more prevalent toward higher redshift is qualitatively consistent with these observational trends. Although our analysis does not directly model gas physics or AGN fueling, it suggests that hierarchical satellite--satellite interactions within recently accreted groups may represent a plausible dynamical channel that facilitates merger-driven gas inflows and thereby contributes to the triggering of IR-bright AGN.}

\rev{Beyond galaxy mergers, these conclusions may also extend to other systems governed by similar dynamical processes, such as globular clusters. As globular clusters are of much lower masses compared to their host satellite or host halo, a direct application of our group-level sinking rate statistics would predict that they barely merge with their hosts. If any of them do merge, a direct extrapolation of the pathway statistics implies that the majority should merge with their host satellites rather than the central galaxy, if they are accreted from other halos. Nevertheless, more realistic simulations of globular cluster formation and evolution would be required to verify these extrapolations.}

\rev{For the Milky Way, observations have already revealed the existence of satellite systems around the Large Magellanic Cloud (LMC)~\cite{sat_of_sat1,sat_of_sat2}. Our measurements of the group-level sinking rate suggest that the direct merger rate onto the LMC should be a factor of $\sim 4$ lower than that onto the Milky Way, while at redshift 1 and above the two merger rates were comparable. Quantitatively, 
integrating the halo-level ($\ell=1$) sinking rate for events with a peak mass ratio of $\mu_{\rm peak}>0.1$ over the redshift interval $0 \leq z < 1$, 
yields an expected average of approximately 0.1 satellite mergers onto the central galaxy per Milky Way-like halo during this period.}


\section{Summary and Conclusions}
\label{sec:summary}
\rev{In this study, we systematically investigate the statistics of subhalo mergers by leveraging the \hbt subhalo finder that defines subhalo mergers as phase-space coalescence, referred to as subhalo ``sinking'' events. Such events unambiguously capture the transition from a decaying to a stalled phase in the orbital evolution of a subhalo, providing a physically meaningful yet numerically robust identification of mergers. 
Our main findings are as follows}:

\begin{enumerate}

    \item \textbf{Most subhalos sink into their direct parents, but cross--level pathways exist.}
    In Figure~\ref{fig:sink_parent_level}, we find that deep-level subhalos predominantly sink into their direct parents. However, tidal stripping, \rev{group accretion}, and limited numerical resolution all contribute to several cross--level sinking pathways.

    \item \textbf{Resolved \rev{sinking events} are predominantly major mergers.}
    Figure~\ref{fig:tot_sink_group} shows that resolved sinking events are predominantly associated with major mergers, with $\mu_{\rm rel}>0.1$, reflecting the stronger dynamical friction experienced by massive satellites. This preference for major mergers persists across hierarchical levels and redshifts.

    \item \textbf{The merger ratio distribution depends on the dynamical age of the host halo.} The average merger ratio is lower in dynamically younger systems, corresponding to halos at a higher redshift or with a larger mass (Figure~\ref{fig:tot_sink_halo}). This is because subhalos with a lower mass ratio are expected to sink earlier due to their earlier accretion time. Consequently, dynamically older halos are more likely to have exhausted their lower mass ratio mergers.

    \item \textbf{Sinking statistics depend strongly on the hierarchical level.}
    Although deep-level subhalos typically have small peak mass ratios relative to their host halos, they can remain massive relative to their direct parents. As a result, satellite--satellite sinking events contribute comparably to, or even exceed, the central--satellite sinking rate at the low-mass end (Figures~\ref{fig:tot_sink_halo} and \ref{fig:tot_sink_redshift}). Therefore, satellite--satellite mergers are an important component of hierarchical subhalo evolution.

    \item \textbf{Satellite--satellite mergers preferentially occur in the halo outskirts.}
    In contrast to central--satellite mergers, which occur near the halo center, Figure~\ref{fig:spatial_sink_detail} shows that sinking events of the deeper-level subhalos preferentially occur at larger radii.

    \rev{\item \textbf{Simulation resolution mostly affects the spatial resolution of the sinking events while the sinking rate statistics remain robust.} Higher-resolution simulations are able to better localize the onset of the events. On the other hand, the time resolution of the sinking events is insensitive to simulation resolution, due to the extremely rapid orbital decay prior to sinking. As a result, the global sinking rate statistics are robust against resolutions.}

    \rev{\item \textbf{Reverse-sinking candidates mainly trace tidally heated satellites.}
    A host subhalo may also sink into the core of its satellite subhalo to produce a ``reverse'' sinking event. These candidates are not ordinary orbital-decay sinking events: they arise when the child subhalo is tidally heated and expands into a diffuse component around a still compact parent core. The child typically becomes unresolved within $1$--$2$ dynamical times, so excluding these transient candidates has little impact on the final subhalo statistics. Therefore, it is reasonable that the current {\hbt} implementation does not explicitly account for this evolutionary channel.}
    

\end{enumerate}


These results establish a coherent picture in which the sinking events in cosmological simulations are governed jointly by dynamical friction, hierarchical accretion, and numerical resolution, providing valuable insights for interpreting the evolution of satellite systems and for informing semi-analytic galaxy formation models with physically motivated multi-level sinking pathways.

\rev{The present analysis parameterizes sinking statistics primarily by host-halo peak mass; future work can extend this framework using pair properties at first accretion into the direct parent. 
Nevertheless, the current analysis offers valuable guidance for developing physically motivated, resolution-independent models of hierarchical mergers among subhalos and satellite galaxies. The data products supporting this work will be made available on Zenodo upon publication: \url{https://doi.org/10.5281/zenodo.21139446}.}

\Acknowledgements{This work is supported by National Key R\&D Program of China (2023YFA1607800, 2023YFA1607801), NSFC(12595312,12595311,12133006), 111 project (No.\ B20019), and the science research grants from the China Manned Space Project (No.\ CMS-CSST-2025-A04). We acknowledge the sponsorship from the Yangyang Development Fund. This project is supported in part by Office of Science and Technology, Shanghai Municipal Government (grant Nos. 24DX1400100, ZJ2023-ZD-001).
ZZ.L. acknowledges the Fundamental Research Funds for the Central Universities (KG202502).
VJFM acknowledges support by NWO through the Dark Universe Science Collaboration (OCENW.XL21.XL21.025). The computation of this work is done on the \textsc{Gravity} supercomputer at the Department of Astronomy, Shanghai Jiao Tong University. }

\InterestConflict{The authors declare that they have no conflict of interest.}

\bibliographystyle{scpma}
\bibliography{sample701}{}

\end{document}